\documentclass{jpp}
\usepackage{graphicx,bm,natbib,color,url}
\usepackage{amsfonts}
\graphicspath{{./fig/}}

\newcommand{\EQ}{\begin{equation}}
\newcommand{\EN}{\end{equation}}
\newcommand{\EQA}{\begin{eqnarray}}
\newcommand{\ENA}{\end{eqnarray}}
\newcommand{\eq}[1]{(\ref{#1})}

\newcommand{\Eq}[1]{equation~(\ref{#1})}
\newcommand{\Eqs}[2]{equations~(\ref{#1}) and~(\ref{#2})}

\newcommand{\Sec}[1]{\S\,\ref{#1}}

\newcommand{\Fig}[1]{figure~\ref{#1}}

\newcommand{\Figs}[2]{figures~\ref{#1} and \ref{#2}}
\newcommand{\Tab}[1]{table~\ref{#1}}

\newcommand{\bra}[1]{\langle #1\rangle}

\newcommand{\meanEMF}{\overline{\mbox{\boldmath ${\cal E}$}} {}}
\newcommand{\tildeEMF}{\tilde{\mbox{\boldmath ${\cal E}$}} {}}

\newcommand{\meanA}{\overline{A}}
\newcommand{\meanB}{\overline{B}}

\newcommand{\meanJ}{\overline{J}}

\newcommand{\tildeemf}{\tilde{\cal E}}

\newcommand{\tildeaa}{\tilde{\bm{a}}}
\newcommand{\tildebb}{\tilde{\bm{b}}}
\newcommand{\meanAA}{\overline{\bm{A}}}
\newcommand{\meanBB}{\overline{\bm{B}}}
\newcommand{\meanJJ}{\overline{\bm{J}}}

\newcommand{\meanUU}{\overline{\bm{U}}}

\newcommand{\hatBB}{\hat{\mbox{\boldmath $B$}}}

\newcommand{\bs}{\boldsymbol}
\newcommand{\zz}{\hat{\mbox{\boldmath $z$}} {}}

\newcommand{\xx}{\mbox{\boldmath $x$} {}}

\newcommand{\uu}{\mbox{\boldmath $u$} {}}

\newcommand{\UU}{\mbox{\boldmath $U$} {}}

\newcommand{\bb}{\mbox{\boldmath $b$} {}}

\newcommand{\BB}{\mbox{\boldmath $B$} {}}

\newcommand{\AAA}{\mbox{\boldmath $A$} {}}
\newcommand{\aaaa}{\mbox{\boldmath $a$} {}}

\newcommand{\JJ}{\mbox{\boldmath $J$} {}}

\newcommand{\nab}{\mbox{\boldmath $\nabla$} {}}
\newcommand{\OO}{\mbox{\boldmath $\Omega$} {}}

\newcommand{\ddelta}{\mbox{\boldmath $\delta$} {}}

\newcommand{\emf}{\mbox{\boldmath ${\cal E}$} {}}

\def\urms{u_{\rm rms}}
\def\Urms{U_{\rm rms}}
\def\Brms{B_{\rm rms}}
\def\meanBrms{\overline{B}_{\rm rms}}

\newcommand{\ii}{{\rm i}}

\newcommand{\dd}{{\rm d} {}}

\def\la{\mathrel{\mathchoice {\vcenter{\offinterlineskip\halign{\hfil
$\displaystyle##$\hfil\cr<\cr\sim\cr}}}
{\vcenter{\offinterlineskip\halign{\hfil$\textstyle##$\hfil\cr<\cr\sim\cr}}}
{\vcenter{\offinterlineskip\halign{\hfil$\scriptstyle##$\hfil\cr<\cr\sim\cr}}}
{\vcenter{\offinterlineskip\halign{\hfil$\scriptscriptstyle##$\hfil\cr<\cr\sim\cr}}}}}

\def\Rm{R_{\rm m}}
\def\Rmc{R_{\rm m}^{\rm crit}}
\def\Imag{\mbox{\rm Im}}
\def\Rey{\mbox{\rm Re}}

\def\kf{k_{\rm f}}

\def\etat{\eta_{\rm t}}

\newcommand{\yan}[5]{~ #1~ #5. {\em Astron. Nachr. }{\bf #2}, #3--#4.}

\newcommand{\yana}[5]{~ #1~ #5. {\em Astron. Astrophys. }{\bf #2}, #3--#4.}
\newcommand{\yanaN}[4]{~ #1~ #4. {\em Astron. Astrophys. }{\bf #2}, #3.}

\newcommand{\ymn}[5]{~ #1~ #5. {\em Month. Not. Roy. Astron. Soc. }
{\bf #2}, #3--#4.}
\newcommand{\ymnS}[5]{~ #1~ #5 {\em Month. Not. Roy. Astron. Soc. }
{\bf #2}, #3--#4.}
\newcommand{\ynat}[5]{~ #1~ #5. {\em Nature }{\bf #2}, #3--#4.}

\newcommand{\yjfm}[5]{~ #1~ #5. {\em J. Fluid Mech. }{\bf #2}, #3--#4.}

\newcommand{\yprl}[5]{~ #1~ #5. {\em Phys. Rev. Lett. }{\bf #2}, #3--#4.}
\newcommand{\yprlN}[4]{~ #1~ #4. {\em Phys. Rev. Lett. }{\bf #2}, #3.}

\newcommand{\yprsa}[5]{~ #1~ #5. {\em Proc. Roy. Soc. Lond. A }{\bf #2}, #3--#4.}
\newcommand{\yptrsa}[5]{~ #1~ #5. {\em Phil. Trans. Roy. Soc. A}{\bf #2}, #3--#4.}
\newcommand{\yjcp}[5]{~ #1~ #5. {\em J. Comp. Phys. }{\bf #2}, #3--#4.}

\newcommand{\ypepi}[5]{~ #1~ #5. {\em Phys. Earth Planet. Int. }{\bf #2}, #3--#4.}
\newcommand{\yapj}[5]{~ #1~ #5. {\em Astrophys. J. }{\bf #2}, #3--#4.}
\newcommand{\yapjN}[4]{~ #1~ #4. {\em Astrophys. J. }{\bf #2}, #3.}

\newcommand{\ypf}[5]{~ #1~ #5. {\em Phys. Fluids }{\bf #2}, #3--#4.}

\newcommand{\ygafd}[5]{~ #1~ #5. {\em Geophys. Astrophys. Fluid Dyn. }{\bf #2}, #3--#4.}
\newcommand{\yphy}[5]{~ #1~ #5. {\em Physica } {\bf #2}, #3--#4.}

\newcommand{\yjour}[6]{~ #1~ #6. {\em #2} {\bf #3}, #4--#5.}

\newcommand{\yjourN}[5]{~ #1~ #5. {\em #2} {\bf #3}, #4.}
\newcommand{\yjourNN}[4]{~ #1~ #4. {\em #2}, #3.}

\newcommand{\ybook}[3]{~ #1~ {\em #2}. #3.}
\newcommand{\ybookS}[3]{~ #1~ {\em #2} #3.}

\newcommand{\ypreN}[4]{~ #1~ #4. {\em Phys.\ Rev. E }{\bf #2}, #3.}

\newcommand{\etal}{{\em et al.}}
\def\half{{\textstyle{1\over2}}}

\title[Mean-field generation in optimal dynamos]{
The nature of mean-field generation in three classes of optimal dynamos
}

\author[Axel Brandenburg and Long Chen]{
Axel Brandenburg$^{1,2,3}$\thanks{
Email address for correspondence: brandenb@nordita.org}
and Long Chen$^{4}$}

\affiliation{
$^1$Nordita, KTH Royal Institute of Technology and Stockholm University, and\\
Department of Astronomy, Stockholm University, SE-10691 Stockholm, Sweden\\
$^2$Laboratory for Atmospheric and Space Physics and JILA,\\
University of Colorado, Boulder, CO 80303, USA\\
$^3$McWilliams Center for Cosmology \& Department of Physics,\\
Carnegie Mellon University, Pittsburgh, PA 15213, USA\\
$^4$Department of Mathematical Sciences, Durham University, Durham, DH1 3LE, UK
}

\date{\today,~ $ $Revision: 1.163 $ \!\! $}

\begin{document}

\maketitle

\begin{abstract}
In recent years, several optimal dynamos have been discovered.
They minimize the magnetic energy dissipation or,
equivalently, maximize the growth rate at a fixed
magnetic Reynolds number.
In the optimal dynamo of Willis (2012, Phys.\ Rev.\ Lett.\ 109, 251101),
we find mean-field dynamo action for planar averages.
One component of the magnetic field grows exponentially while the other
decays in an oscillatory fashion near onset.
This behavior is different from that of an $\alpha^2$ dynamo, where the
two non-vanishing components of the planar averages are coupled and have
the same growth rate.
For the Willis dynamo, we find that the mean field is excited
by a negative turbulent magnetic diffusivity, which
has a non-uniform spatial profile near onset.
The temporal oscillations in the decaying component are caused
by the corresponding component of the diffusivity tensor being
complex when the mean field is decaying and, in this way, time-dependent.
The growing mean field can be modeled by a negative magnetic diffusivity
combined with a positive magnetic hyperdiffusivity.
In two other classes of optimal dynamos of Chen {\em et al.}
(2015, J.\ Fluid Mech.\ 783, 23), we find, to some extent, similar
mean-field dynamo actions.
When the magnetic boundary conditions are mixed,
the two components of the planar averaged field grow at different rates
when the dynamo is 15\% supercritical. 
When the mean magnetic field satisfies homogeneous boundary conditions
(where the magnetic field is tangential to the boundary),
mean-field dynamo action is found for one-dimensional averages, 
but not for planar averages.
Despite having different spatial profiles, 
both dynamos show negative turbulent magnetic diffusivities.
Our finding suggests negative turbulent magnetic diffusivities 
may support a broader class of dynamos than previously thought, 
including these three optimal dynamos. 
\end{abstract}

\section{Introduction}

Since the works of Varley, Wheatstone, and Siemens of around 1867,
we know that electromagnetic dynamos can be self-excited, i.e., they work
without permanent magnets to turn kinetic energy into electromagnetic energy.
Unlike those technical dynamos with wires, homogeneous dynamos work
in uniformly conducting media \citep{Lar19}.
They are prone to short-circuiting themselves, so for a long time it was
unclear whether they can work at all.
Indeed, it became clear that axisymmetric magnetic fields cannot be
sustained by a dynamo \citep{Cow33}. 
Axisymmetric flows, on the other hand, are capable of producing dynamos,
but the resulting magnetic field is necessarily nonaxisymmetric
\citep{Gai70,Pon73,DJ89}.
\cite{Mos90} found that the critical magnetic Reynolds
number, i.e., the ratio of inertial to resistive electromagnetic forces,
is rather large for the dynamo of \cite{Gai70} to be excited.
This means that the typical scale and velocity can be very large,
so an experimental verification is difficult for that flow.
For the Ponomarenko dynamo, by contrast, the critical magnetic Reynolds
number is sufficiently low so that an experimental verification
was successful \citep{Riga1}.

Another self-excited dynamo arrangement that has been subjected to
experimental verification is that of \cite{Her58}.
The critical magnetic Reynolds number is again very large, but by
using solid copper rotors that are in electric contact within a large
copper block, it was possible to reach supercritical conditions
\citep{LW63,LW68}.
Modeling the Herzenberg dynamo numerically has been possible by using
relatively large rotors that are close together \citep{BMS98}.
From a numerical point of view, however, it is more advantageous to
use periodic flow patterns.
For example the Roberts flow I in a cubic domain has a critical magnetic
Reynolds number of about 5 based on the root mean square (rms) velocity
and the wavenumber of the flow.
By comparison, the critical magnetic Reynolds number for the ABC flow
\citep{GF86} is about 15.

Indeed, certain periodic flows are better than others at producing dynamos.
The optimal flow of ABC type, for example, was identified by \cite{A11}. 
Other studies \citep{PK10,PWK12} on pipe flows led \cite{W12} to consider
this as a variational problem.
The optimal steady flow to excite a kinematic dynamo is found iteratively. 
Due to the arbitrarily low kinetic energy required to excite a dynamo \citep{P15}, 
the magnetic Reynolds number should be based on the rms vorticity rather than the rms speed,
which we refer to as $R_\omega$ in this paper.
A series of optimization studies followed.
\cite{CHJ15} considered physical boundary conditions, and \cite{CHLLLJ18} applied the same method to a sphere. 
The optimal axisymmetric flow in a sphere has also been identified in the thesis of \cite{C18}. 
All optimal flows have lowered the critical $R_{\omega}$ significantly. 

It is often thought that some kind of swirl or helicity in the flow is
important, but this is not generally true \citep{GFP88}.
Three of the four flows studied by \cite{Rob72} have no helicity and
yet they can produce mean magnetic fields, as defined by planar averaging.
In fact, different types of flow geometries can produce very different
types of dynamos: small-scale dynamos, large-scale dynamos, those with
an $\alpha$ effect and those without, etc. 
Nonhelical isotropic turbulence only leads to small-scale dynamo
action when the magnetic Reynolds number based on the rms speed exceeds
a critical value that is between 40 and 200, 
depending on the magnetic Prandtl number---the ratio
of kinematic viscosity to magnetic diffusivity \citep{Isk07,Bra11}.
Those dynamos produce magnetic fields at scales as small as the resistive
scale---the smallest scale where turbulent magnetic fields exist.

Some flows also act as large-scale dynamos, which produce magnetic fields
that can be detected even after spatial averaging over certain directions.
Such dynamos are also referred to as mean-field dynamos.
Here we show that some of the optimal dynamos do indeed produce mean fields.
We also examine the nature of such mean field dynamo action.

\section{Instructive examples of mean-field dynamos}

A particularly famous example of a large-scale dynamo is the Roberts flow~I,
a flow with maximum kinetic helicity. The flow is of the form
$\UU=\kf\psi\zz+\nab\times(\psi\zz)$, where $\psi=\cos kx\cos ky$
\citep{Rob72}.
The critical magnetic Reynolds number, based on the rms velocity and
the effective wavenumber $\sqrt{2}k$, is 3.9 for
a domain of length $L=2\pi/k$ in the $z$ direction \citep{BS05}.
Interestingly, a flow of the form $\UU=\kf\psi\zz+\nab\times(\phi\zz)$,
where $\phi=\sin kx\sin ky$ is phase-shifted in the $x$ and $y$ directions
by $\pi/2$ relative to $\psi$ (Roberts flow II), has zero pointwise
kinetic helicity.
So, there is no swirl whatsoever, and yet, it produces not only a
magnetic field, but one with non-vanishing $xy$ averages, although it
requires $L\ge2\pi/(0.64\,k)\approx9.8/k$ \citep{RDRB14}.

In the following examples, we focus on dynamos that produce a non-vanishing
mean field obtained by averaging over the $xy$ plane of size $L^2$.
This mean field is therefore defined as $\meanBB=\int\BB\,\dd x\,\dd y/L^2$.
In the case of Roberts flow~I, the resulting mean field is of the form
$\meanBB=B_0(t)(\sin (kz+\varphi),\cos (kz+\varphi),0)$, where $\varphi$ is an         
arbitrary phase, $k$ is the wavenumber, and $B_0(t)$ is a time-dependent amplitude.
In the case of Roberts flow~IV, the velocity has no net, but still pointwise
helicity. The mean magnetic field is time-dependent and its $x$ and $y$ 
components evolve independently of each other, i.e., 
\EQ
\meanBB(z,t)=(\meanB_x, \meanB_y, 0)=\Big(B_{0x}(t)\cos (k^{(x)}z+\varphi_x),\; B_{0y}(t)\cos (k^{(y)}z+\varphi_y),\; 0\Big),
\EN
where $B_{0x}(t)$ and $B_{0y}(t)$ are time-dependent amplitudes, $\varphi_x$
and $\varphi_y$ are phases, and $k^{(x)}$ and $k^{(y)}$ are wavenumbers
for the $x$ and $y$ components, respectively.
In principle, 
the values of $\meanB_x$ and $\meanB_y$ for a mean-field dynamo can 
be different from each other.
For example, if $B_{0y}=0$, the magnetic field can just be
$\meanBB=(B_{0x}\cos kz,0,0)$, i.e., with only one component.
In fact, the Robert flows II--IV all produce dynamos where $\meanB_x$ and
$\meanB_y$ evolve independently of each other, albeit at the same rate,
i.e., $\dd\ln B_{0x}/\dd t=\dd\ln B_{0y}/\dd t$; see
\cite{DBM13,RDRB14}.
We are not aware of any earlier demonstration of a case
where the growth rates of different components are not the same.

Solutions to large-scale or mean-field dynamos can be obtained if
the mean electromotive force can be expressed in terms of the mean field.
Here, the mean electromotive force is defined as
$\meanEMF=\overline{\uu\times\bb}$, where overbars denote $xy$ averaging
and lowercase symbols denote fluctuations around the mean field, i.e.,
$\uu=\UU-\meanUU$ and $\bb=\BB-\meanBB$ are the fluctuating velocity
and magnetic fields.
The general relationship between $\meanBB$ and $\meanEMF$ is
in terms of a convolution of the form
\EQ
\meanEMF=\int\!\!\!\!\int K_{ij}(z-z',t-t')\,\meanB_j(z',t')\,\dd z'\,\dd t',
\EN
where $K$ is an integral kernel. 
In the case of Roberts flow~I, when the mean magnetic field is marginally excited,
the kernel is approximately of the form
\EQ
K_{ij}(z,z',t,t')\approx\delta(z-z')\delta(t-t')
(\alpha\delta_{ij}-\etat\epsilon_{i3j}\partial/\partial z'),
\label{kernelalpeta}
\EN
where $\alpha=(\eta+\etat)k$ in the marginally excited case.
In the case of Roberts flow~II, the components of $K_{ij}$ cannot
be described by an instantaneous relationship, but there is a
turbulent pumping effect with a certain time delay \citep{RDRB14}.
Thus, the $x$ and $y$ components evolve independently of each other.
The dynamo for Roberts flow~III is similar to that for Roberts flow~II,
except that the $x$ and $y$ components of the mean magnetic field
experience pumping velocities that point in opposite directions.

Finally, Roberts flow~IV is again given by an equation similar to
\eq{kernelalpeta}, but with $\alpha=0$ and $\etat$ being negative on
the length scales of interest.
At smaller scales, however, $\etat$ is always positive, which is necessary
so as to ensure stability at small length scales.
This can be accounted for by adding a magnetic hyperdiffusivity,
corresponding to an additional third-order derivative term in
\Eq{kernelalpeta}.
We return to this at the end of the paper.
With these preparations in place, we are now in a position to characterize
the dynamos driven by the aforementioned optimized flows.

In this study, we encounter examples of dynamos that share
similarities with some of the cases discussed above.
In particular, we find cases that do exhibit this type of
unusual behavior with two components evolving independently of each other.
For example, the Willis dynamo is even more bizarre than that of
Roberts flow~IV, because the two horizontal components of $\meanBB$
evolve differently, with growth rates that have even different signs.
Robert flow~I, by contrast, is maximally helical and leads to an
$\alpha$ effect that couples the two horizontal components of $\meanBB$.
Those dynamos are called $\alpha^2$ dynamos, because the $\alpha$ effect
is responsible for producing $\meanB_x$ from $\meanB_y$ and
for producing $\meanB_y$ from $\meanB_x$.
We talk about $\alpha\Omega$ dynamos when shear in (say) the $y$
direction ($\Omega$ effect) is responsible for producing $\meanB_y$
from $\meanB_x$.
The flows of \cite{W12} and \cite{CHJ15} turn out {\em not} to be of that type.
Below, we present a more detailed analysis of these optimal dynamos.

\section{Essentials of dynamos driven by optimized flow}
\label{Essentials}

The goal of this study is to determine the nature of dynamo action driven by 
the flows of \cite{W12} and \cite{CHJ15}.
We discuss three types of optimal flows 
with distinct boundary conditions for their excited magnetic eigenmodes, 
referred to as Willis, NNT, and TTT cases.
We begin by briefly explaining the essential technique to obtaining
these optimal flows, then describe the dominant Fourier modes of the
dynamo solution, if there is any.

Basically, the optimization method belongs to the family of constrained
optimizations.
Given a Lagrangian as follows, 
\EQA\label{lagrangian}
\mathcal{L}&=& \ln \left\langle  \bs{B}_T^2  \right\rangle \nonumber  \\
&&-\lambda_1\left(\langle  \bs{\omega}^2\rangle-1\right)  - \lambda_2 \left(\langle  \bs{B}_0^2\rangle-1\right)
-\left\langle  \Pi_1\,   \nabla \cdot \bs{U}\right\rangle -  \langle\Pi_2 \nabla \cdot \bs{B}_0\rangle \nonumber  \\
&& -\int_0^T\left\langle  \bs{B}^{\dagger} \cdot \left[ \partial_t \bs{B} -\nabla \times (\UU \times \bs{B}) - R_\omega^{-1} \nabla^2 \bs{B}  \right]\right \rangle \textrm{d} t,
\ENA
where $\bs{\omega}=\nabla\times\bs{U}(\bs{x})$, $\bs{B}_T=\bs{B}(\bs{x},T)$,
$\langle\cdots\rangle=V^{-1}\int \cdots \dd V$ denotes the volume average,
and $T$ is a fixed time that is long enough to filter out the transient growth. 
The first term in \Eq{lagrangian} is a proxy of the growth rate that we want to
maximize, whilst the other terms are constraints from a kinematic dynamo model
(the backreaction from the magnetic field on the flow is not considered). 
We search for the optimal flow $\bs{U}$ and the initial field $\bs{B}_0$
such that all variations vanish:
\EQ
\delta \mathcal{L}(\bs{U}, \bs{B}, \bs{B}^\dagger, \bs{B}_0, \bs{B}_T,\lambda_1,\lambda_2,\Pi_1,\Pi_2 )=0.
\EN
Setting $\delta\mathcal{L}/\delta\bs{B}^\dagger=0$ gives the induction equation
\EQ\label{foreq}
\partial_t \bs{B} =\nabla \times (\UU \times \bs{B}) + R_\omega^{-1} \nabla^2 \bs{B},
\EN
and setting
$\delta\mathcal{L}/\delta\bs{B}=0$ gives the adjoint induction equation,
\EQ\label{adjeq}
\partial_t \bs{B}^{\dagger} = \UU\times(\nabla \times \bs{B}^{\dagger}) - R_\omega^{-1} \nabla^2  \bs{B}^{\dagger}.
\EN
The optimization procedure is based on the two equations above.
Starting from some fields $\bs{U}$ and $\bs{B}_0$, we first evolve the
system forward in time using \Eq{foreq} until time $T$, then backward in
time using \Eq{adjeq}, and finally we use $\delta\mathcal{L}/\delta\bs{U}$
and $\delta\mathcal{L}/\delta\bs{B}_0$ as gradients to update $\bs{U}$
and $\bs{B}_0$.
A detailed optimization algorithm is described in \cite{CHJ15}.

The resulting optimal flow $\bs{U}$ and 
the corresponding least decaying magnetic eigenmode 
show drastically different features for the three cases we are interested in.
The Willis case represents the most efficient solution (has lowest critical $R_{\omega}$) 
with periodic boundary conditions for the flow and the magnetic eigenmode. 
Since the optimization algorithm does not fix the orientation of fields,
any transformation such as shift ($T_\delta$), rotation, or reflection ($\mathcal{R}$) gives
the same equivalent optimal flows,
\EQ
\UU(\bs{x})=\mathcal{R}^{-1}\widetilde{\UU}(\tilde{\bs{x}}(\bs{x})),
\EN
where $\tilde{\bs{x}}(\bs{x})=\mathcal{R}\bs{x}+ T_\delta$ gives the relation of two coordinates. 
Here, tildes denote quantities in the original coordinate.
The Willis flow can be approximately described by a large-scale dominant
flow (up to $1\%$ difference in the vorticity norm $\Vert\omega\Vert_2$)      
in the original coordinate as:
\EQ
\label{willis*}
\frac{\widetilde{\UU}}{{\tilde{U}}_{\rm rms}}\approx\frac{2}{\sqrt{3}}\big(\sin \tilde{y} \cos \tilde{z},\, \sin \tilde{z} \cos \tilde{x}\
,\, \sin \tilde{x} \cos \tilde{y}\big).
\EN
For the flow we use, the actual form in the transformed coordinate is 
\EQ \label{willist}
\frac{\UU}{\Urms}\approx\frac{2}{\sqrt{3}}\left(\sin \left(z+\pi/2\right) \cos \left(y-\pi/4\right),\, \sin x \cos \left(z+\pi/2\right),\, \sin(y-\pi/4) \cos x\right).
\EN
The two sets of coordinates are related by
$(\tilde{x},\tilde{y},\tilde{z})=(x, z +\pi/2, y-\pi/4)$.
The magnetic eigenmode of this dynamo can also be approximated by
a simple field (up to $2\%$ difference in energy) as
\EQ \label{willisbt}
\bs{B}_T\approx \left [\begin{array}{cc}
 ~~0.138\, \cos z +0.810 \, \sin z\\
 -0.802  \, \cos x + 0.179\, \sin  x\\
 -0.538 \, \cos y-0.622\, \sin  y
 \end{array} \right ],
\EN
in the transformed coordinate.
The $x, y, z$ components of \Eq{willisbt} each vary only in one direction.
If small fluctuations are added to this eigenmode, 
we would expect to find a non-zero mean field when taking
any of the planar averages. 

In \cite{CHJ15}, a combination of sine and cosine functions 
is used to mimic physical boundary conditions in a domain of size 1.
The flow satisfies non-penetrating boundary conditions and therefore only
sine functions are allowed in the direction perpendicular to the boundary.
In this study, we rescale the flows of \cite{CHJ15} such that they become
periodic in a $(2\pi)^3$ domain while retaining the boundary conditions
in a $\pi^3$ domain.
In the extended $(2\pi)^3$ domain, the general form is given by
\EQ
\label{opflow}
\UU   = \!\! \sum_{m_x,m_y,m_z \in \mathbb{N}} \left [\begin{array}{cc} 
 a_x (m_x,m_y,m_z) \, \sin (m_x x) \, \cos (m_y y) \, \cos (m_z z) \\
 a_y (m_x,m_y,m_z) \, \cos (m_x x) \, \sin (m_y y) \, \cos (m_z z) \\
 a_z (m_x,m_y,m_z) \, \cos (m_x x) \, \cos (m_y y) \, \sin (m_z z)
 \end{array} \right ],   \label{eq:Uexpand}
\EN
and $a_i(m_x,m_y,m_z)$ with $i=x,y,z$ being the spectral coefficients.
This is similar to the Taylor--Green flow
but with the sine and cosine functions swapped. 
Meanwhile, 
the magnetic boundary conditions can be either superconducting or pseudo-vacuum.
In terms of spectral representations, this means either using sine functions
(T) or cosine functions (N) in the direction perpendicular to the boundary.
The corresponding magnetic eigenmodes then have four possible combinations:
NNT, NTT, NNN, and TTT in $x,y,z$ respectively. 
The NNT/NTT and NNN/TTT pairs give the same dynamo up to a symmetry
transformation \citep{FP13}. 
Thus, only one solution from each pair is chosen for this study. 
The NNT magnetic field has the general form
\EQ
\BB'_{\rm NNT}=\!\!\sum_{m_x,m_y,m_z \in \mathbb{N}} \left [\begin{array}{cc}
a_x (m_x,m_y,m_z) \, \cos (m_x x) \, \sin (m_y y) \, \cos (m_z z) \\
a_y (m_x,m_y,m_z) \, \sin (m_x x) \, \cos (m_y y) \, \cos (m_z z) \\
a_z (m_x,m_y,m_z) \, \sin (m_x x) \, \sin (m_y y) \, \sin (m_z z)
 \end{array} \right ],
\EN
which corresponds to normal field boundary conditions
in the $x$ and $y$ directions, and perfectly conducting
boundaries in the $z$ direction. 
$\BB'_{\rm TTT}$ has the same general representation as \Eq{eq:Uexpand}.
This is because the boundary conditions for both fields forbid normal
components across the boundary, but allow tangential components.

To distinguish the two cases,
the optimal flows are named after the boundary conditions of their corresponding magnetic eigenmodes.
For up to $83\%$ of the total enstrophy, $\langle\bs{\omega}^2\rangle$, the optimal NNT flow can be approximated
by the velocity $\UU\approx\nabla\times\bs{\psi}$ with
\EQ
\quad\left [\begin{array}{c} 
 \psi_x(x,y,z) \\
 \psi_y (x,y,z)\\
 \psi_z (x,y,z) \end{array} \right ]   =  \left [\begin{array}{cc} ~~0.16& \sin 2 y \, \sin2z \\
~~ 0.77 & \quad \sin x\, \sin z \\
-0.18 &  \sin 2x \ \sin 2y \\
 \end{array} \right ].
\label{eq:vecpnnt}
\EN
The leading components of the magnetic eigenmode are given in \cite{CHJ15}. 
In particular, the most energetic Fourier mode takes $39\%$ of the total energy,
and varies only in one direction:
\EQ
{B'_x}_{\rm NNT}(0,1,0)= 0.883\, \sin y,
\EN
where $(0,1,0)$ are the wavenumbers in $x, y, z$ respectively. 
We expect to see some contribution from $B'_x$ for the $xz$ averages 
taken in the $(2\pi)^3$ domain.  
The other dominant Fourier components depend on at least two directions,
hence have zero planar averages. 

The TTT case has neither a dominant flow 
nor a dominant magnetic field that varies only in one direction. 
The optimal TTT flow is highly localized, but 
has approximately equal enstrophy per direction ($\langle\omega_i^2\rangle, i=x,y,z$);
see also \cite{CHJ15} for the streamline plot. 
We do not expect to find a planar mean field near onset for this type of dynamo.

With three types of boundary conditions, 
we get three optimal solutions:
the Willis and NNT dynamos both have simple large-scale flows, 
except the NNT dynamo breaks the symmetry in one direction,
and the TTT dynamo has a complex localized flow.
These three dynamo solutions were computed without 
imposing specific physical properties of the flow \textit{a priori}. 
While this approach allows us to remove bias and explore 
the full parameter space of solutions, it leaves the question of how to 
interpret the dynamo action. 
In the next section, we discuss how to 
extend the analysis of dynamos using mean-field theories.

\section{Mean-field approaches to analyzing the dynamos}
\label{analyze_dynamo}

To characterize the nature of dynamo action for the
three optimized flow problems, we use both direct numerical
simulations (DNS) and the test-field method (TFM).
The DNS refer to the numerical solution of \Eq{foreq}, whereas the TFM
refers to solutions of the evolution equations for the fluctuations around
a given planar averaged mean field, which is one of four test fields, $\meanBB^T$.
The TFM allows us to extract turbulent transport coefficients from the
underlying flow fields \citep{Sch05,Sch07}; see also \cite{BCDSHKR10}
for a review.
We use numerical representations of the flows of \cite{CHJ15}
for NNT and TTT dynamos in their original form,
i.e., the rms of velocity is unnormalized, 
but extend the domain to $(2\pi)^2$ to match with the periodic boundaries,
and also the reproduced Willis dynamo using a similar algorithm.
The corresponding data files for these three flow fields can be found
in the online material \citep{BC19} for the published data sets used to
compute each of the figures of the present paper.
For comparison, we also use the original formulation of \cite{W12},
denoted by Willis$^*$ (with an asterisk), which is written as
$\UU=(2/\sqrt{3})(\sin y\cos z,\sin z\cos x,\sin x\cos y)$
in the same coordinate as other flows, as opposed to the transformed form in \Eq{willist}.
All these flows and the resulting magnetic fields are periodic in the
$(2\pi)^3$ volume, but for the NNT and TTT cases, we also select a $\pi^3$
subvolume with the appropriate boundary conditions.
The flows are represented on a $32^3$ mesh, except for the TTT case,
which is represented on a $48^3$ mesh.
For the NNT and TTT cases in $\pi^3$ subvolumes,
we select the first $17^3$ and $25^3$ meshpoints,
respectively, which include the points on the boundaries.

\subsection{Characterizing the growth of the mean field}
\label{grow_mean}

The growth of the mean field can be characterized through averaging.
From the kinematic model, we know the spatial distribution of the magnetic eigenmode.
The choice of averaging is then determined by the dominant magnetic field components. 
We take $xy$ averages for the Willis and TTT cases,
and $xz$ averages for NNT case, and denote those by an overbar.
To compute rms values, we employ volume averages, denoted by angle brackets.
The rms values of the velocity, $\Urms\equiv\bra{\UU^2}^{1/2}$,
are listed in \Tab{Tsummary}.
The kinetic helicity integrated over the full domain is zero,
but its local value at arbitrary points in the domain is finite. 
For the NNT flow, the kinetic helicity integrated over the $\pi^3$ domain is also zero.
By contrast, 
the TTT flow lacks the symmetry to have a net zero helicity in the $\pi^3$ domain.

To obtain $\BB=\nab\times\AAA$, we solve for the magnetic vector
potential $\AAA$.
Its evolution is governed by the uncurled induction equation,
\begin{equation}
{\partial\AAA\over\partial t}=\UU\times\BB+\eta\nabla^2\AAA.
\label{dAdt}
\end{equation}
We solve this equation using the {\sc Pencil Code}\footnote{\url{
https://github.com/pencil-code}, DOI:10.5281/zenodo.2315093}, which is a
high-order public domain code for solving partial differential equations,
including the induction equation that is of interest here, as well as
the test-field equations that are discussed in the next section.
The code uses sixth order finite differences in space and the third order
2N-RK3 low storage Runge--Kutta time stepping scheme of \cite{Wil80}.

When $\eta$ is below a certain critical value, $\eta_{\rm crit}$,
the value of the rms magnetic field, $\Brms$, grows exponentially
proportional to $e^{\lambda t}$ if $\lambda$ is a constant.
For oscillatory fields, $\lambda$ can be complex such that
$2\pi/\Imag\lambda$ is the period of the oscillation, and $\Rey\,\lambda$
is the growth rate, which can be estimated from the logarithmic derivative
of $\Brms$ or its envelope for oscillatory fields.
By experimenting with different values of $\eta$, and by interpolation,
we find the critical value $\eta_{\rm crit}$ below which the dynamo
is excited.

\subsection{Quantitative analysis using the TFM}
\label{STFM}

The averaged evolution or mean-field equation reads
\begin{equation}
{\partial\meanAA\over\partial t}=\meanUU\times\meanBB+\meanEMF+\eta\nabla^2\meanAA,
\label{dAmdt}
\end{equation}
where $\meanEMF=\overline{\uu\times\bb}$ is the electromotive force from
the fluctuating velocity and magnetic fields, $\bb=\nab\times\aaaa$ is the
small-scale magnetic field, and $\aaaa$ is a solution to the equation that
results by subtracting \Eq{dAmdt} from \Eq{dAdt}, which yields
\begin{equation}
{\partial\aaaa\over\partial t}
=\meanUU\times\bb+\uu\times\meanBB+\emf'+\eta\nabla^2\aaaa,
\label{dadt}
\end{equation}
where $\emf'=\emf-\meanEMF$ is the fluctuation of the electromotive force.
This equation contains $\meanBB$, but it can also be formulated for
arbitrary mean fields, which we then call test fields, $\meanBB^{\rm T}$.
The goal is to determine solutions to this equation for sufficiently
many independent test fields so that we can assemble all eight unknowns,
$\alpha_{ij}$ and $\tilde{\eta}_{ij}$, for $i,j=1,2$ to the equation
\begin{equation}
\tildeemf_{i}^{\rm T}=\tilde{\alpha}_{ij}\meanB_j^{\rm T}
-\tilde{\eta}_{ij}\meanJ_j^{\rm T},
\label{EMF_TFM}
\end{equation}
where $\meanJJ^{\rm T}=\nab\times\meanBB^{\rm T}$.
In general, the actual mean fields are neither constant in space nor in
time, so we need to sample all Fourier modes in $z$ and $t$
to capture the full dependence.
(Here and in the following, we mean $y$ instead of $z$
when dealing with the NNT flow.)
Fourier transformed variables are denoted by tildes.
The electromotive force is then written in the form
\begin{equation}
\tilde{\cal E}_i(z,t)=\int\left[
\tilde{\alpha}_{ij}(z,t,k,\omega)\tilde{B}_j(k,\omega)
-\tilde{\eta}_{ij}(z,t,k,\omega)\tilde{J}_j(k,\omega)
\right]\dd k\,\dd\omega.
\label{emfizt}
\end{equation}
The $z$ profiles of all components of $\tilde{\alpha}_{ij}(z,t,k,\omega)$
and $\tilde{\eta}_{ij}(z,t,k,\omega)$ are obtained by solving the
test-field equations for different values of $k$ and $\omega$, where $k$
is the wavenumber and $\omega$ is the frequency in Fourier space.

Here, $\tilde{\BB}=\int e^{-\ii(kz-\omega t)}\meanBB\,\dd z\,\dd t$ is the
Fourier transformed mean field, and likewise for $\tilde{\JJ}$.
For determining $\tilde{\alpha}_{ij}$ and $\tilde{\eta}_{ij}$, we solve
four copies of \Eq{dadt}, each with one of four different test fields;
$\meanBB^{\rm T}=(\sin kz,0,0)$, $(\cos kz,0,0)$, $(0,\sin kz,0)$,
and $(0,\cos kz,0)$.
We then solve the test-field equation forward in time. 
Given that $\UU$ is constant in time, 
we can use a relaxation method \cite[see][for details]{RDRB14} and
rewrite the test field as $\meanBB^{\rm T}\to e^{-\ii\omega t}\hatBB^{\rm T}(z;\omega)$.
This notation is not to be confused with the solutions to the adjoint
problem, $\bs{B}^{\dagger}$, discussed in \Sec{Essentials}.

We solve the Fourier-transformed complex
equations for the response to each of the test fields.
Those equations are given by \citep{RDRB14}
\begin{equation}
\ii\omega\tilde{\aaaa}^{\rm T}+\uu\times\hatBB^{\rm T}
+(\uu\times\tilde{\bb}^{\rm T})'+\eta\nabla^2\tilde{\aaaa}^{\rm T}=0,
\label{ComplexTFM}
\end{equation}
where $\tildebb^{\rm T}=\nab\times\tildeaa^{\rm T}$ are the solutions,
tildes denote Fourier transform in time,
$\tildeaa^{\rm T}(\xx,\omega)=\int \aaaa^{\rm T}(\xx,t) e^{-\ii\omega t}\dd t$,
and lowercase letters
and primes denote fluctuations about the planar average.
We then compute the desired transport coefficients in Fourier space,
$\tilde{\alpha}_{ij}$ and $\tilde{\eta}_{ij}$ by measuring
$\tildeEMF^{\rm T}=\overline{\uu\times\tildebb^{\rm T}}$ and then solving
\Eq{EMF_TFM}.
The diagonal components of the sum $\eta+\tilde{\eta}_{ij}$ act as
an {\em effective magnetic diffusivity}, which has to be overcome by the
other inductive effects in order for mean-field dynamo action to occur.
We apply the complex TFM to the three optimal flows discussed earlier.

\section{DNS results and spatial averages for the different dynamos}

\begin{table}\caption{Summary of parameters for various flows.
For the 15\% supercritical cases, the values of $q\equiv\meanBrms/\Brms$
are given along with the corresponding values of $\eta$ (sixth column),
the growth rate $\lambda$ of the rms magnetic field, and
those of the mean-field components $\meanB_x$ and $\meanB_y$
(or $\meanB_z$ for the $xz$ averaged NNT flow), denoted
by $\lambda_{\rm mean}^{(x)}$ and $\lambda_{\rm mean}^{(y,z)}$, respectively.
Their imaginary parts give the frequency of oscillatory field components.
The asterisk on the value 0.45 for $q$ denotes that a columnar
$z$ average has been used in this case (see text).
}\vspace{12pt}\centerline{\begin{tabular}{l|cccc|cclccc}
flow   &  mesh  & $\urms$ & $\eta_{\rm crit}$ & $\Rmc$ & $\eta$ & $\Rm$
& $\;\;q$ & $\lambda$ & $\lambda_{\rm mean}^{(x)}$ & $\lambda_{\rm mean}^{(y,z)}$ \\
\hline
ABC       & $32^3$ & 1.73  & 0.112 & 15.5 & 0.097 & 17.9 &0.18& $0.006$ & $ 0.006+0.62\ii$& $ 0.006+0.62\ii$ \\
Roberts   & $32^3$ &   1   & 0.181 & 5.52 & 0.158 & 6.33 &0.65& $0.029$ & $ 0.029       $ & $ 0.029$  \\
Willis$^*$& $32^3$ &   1   & 0.568 & 1.76 & 0.498 & 2.01 &0.57& $0.068$ & $-0.42+0.80\ii$ & $ 0.068$  \\
Willis    & $32^3$ & 0.705 & 0.403 & 1.75 & 0.350 & 2.01 &0.57& $0.052$ & $ 0.052$        & $-0.26 +0.58\ii$ \\
NNT       & $32^3$ & 0.594 & 0.133 & 4.47 & 0.116 & 5.12 &0.63& $0.014$ & $ 0.014$        & $-0.048\;\;\,$ \\
TTT       & $48^3$ & 0.336 & 0.083 & 4.05 & 0.072 &4.67&$0.45^*$&$0.034$& $ 0.004$   & $ 0.006$  \\
\label{Tsummary}\end{tabular}}\end{table}

The critical magnetic diffusivity, $\eta_{\rm crit}$, 
along with the rms velocities and other relevant
parameters such as the critical magnetic Reynolds numbers,
\begin{equation}
\Rmc=\urms/\eta_{\rm crit}k_1,
\label{rmdef}
\end{equation}
are listed in \Tab{Tsummary}.
Here $k_1$ is an estimate of the relevant wavenumber of the
magnetic field---usually the lowest in the domain.
We use $k_1=1$ in all cases, and fluctuations are periodic.
We also list the ratio $q\equiv\meanBrms/\Brms$ of the rms values
of the mean field, $\meanBrms$, to that of the full field, $\Brms$,
for cases that are about 15\% supercritical.
The Karlsruhe and Riga dynamo experiments were
only about $10\%$ supercritical,
as measured by the magnetic Reynolds number based on the characteristic flow speed \citep{G03,DE07}.
Testing the near-critical response of the mean field is, 
therefore, a useful way of characterizing a weakly supercritical dynamo. 

For comparison, we have included in \Tab{Tsummary}
the results for the more familiar ABC and Roberts flows.
The ABC flow \citep{Chi70} is given by
$\UU=(\sin z+\cos y,\;\sin x+\cos z,\;\sin y+\cos x)$.
The equation for the ABC flow is similar to that for the Willis flow,
except that the multiplications in the latter are replaced by plus signs
in the ABC flow.

\subsection{Dynamo action in the Willis dynamo}

\begin{figure}\begin{center}
\includegraphics[width=0.8\columnwidth]{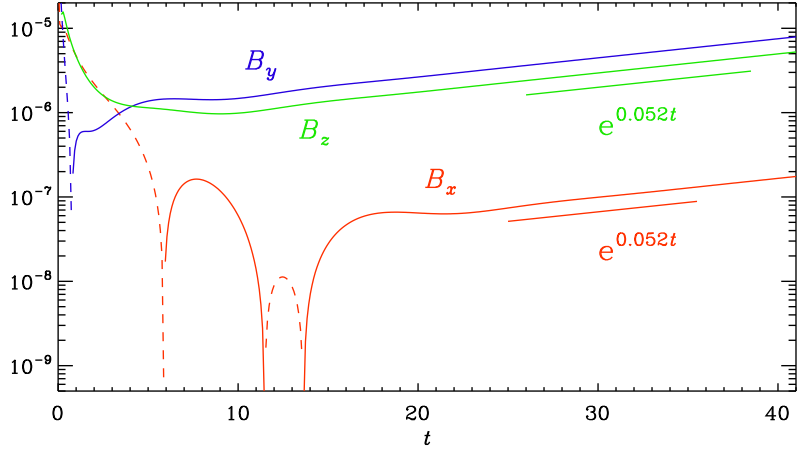}
\end{center}
\caption[]{
The three components of the magnetic field, $B_x$ (red), $B_y$ (blue),
and $B_z$ (green), at an arbitrarily selected point $\xx_\ast$ within the domain
for the Willis flow with $\eta=0.35$, which is supercritical.
All three components begin to grow exponentially at the same rate.
Solid (dashed) lines denote positive (negative) values.
}\label{pts_bxby}
\end{figure}

For the analytically given flow Willis$^*$, which has a root-mean-square
(rms) velocity of unity, we find mean-field dynamo action for
$\eta<\eta_{\rm crit}=0.568$.
This corresponds to $\Rmc=1.76$.
For the numerically optimized flow ``Willis'' (without asterisk), which we
focus on in the rest of this paper, the rms velocity is smaller and
$\eta_{\rm crit}=0.403$.
This corresponds to $\Rmc=1.75$, so it is only slightly easier to excite
than Willis$^\ast$.
The following considerations apply all to the latter, numerically
optimized flow.
In the supercritical case, here using $\eta=0.35$, all three
components of $\BB$ are seen to grow exponentially in time with
$\lambda\approx0.052$; see \Fig{pts_bxby}.

\begin{figure}\begin{center}
\includegraphics[width=0.8\columnwidth]{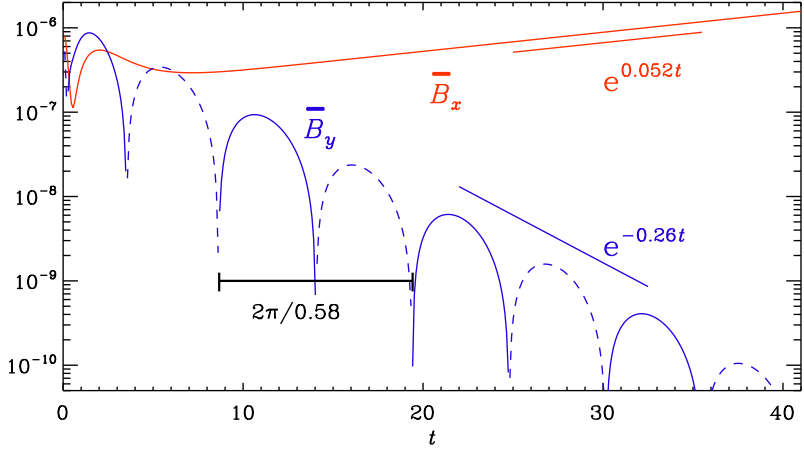}
\end{center}\caption[]{
Evolution of $\meanB_x(z_\ast,t)$ (red) and $\meanB_y(z_\ast,t)$ (blue)
for the Willis flow at a fixed position $z_\ast$.
Note that $\meanB_y$ decays in an oscillatory fashion with the frequency
$\Imag\,\lambda=0.58$.
Again, solid (dashed) lines denote positive (negative) values.
}\label{pxyaver_bxby}
\end{figure}

\begin{figure}\begin{center}
\includegraphics[width=\columnwidth]{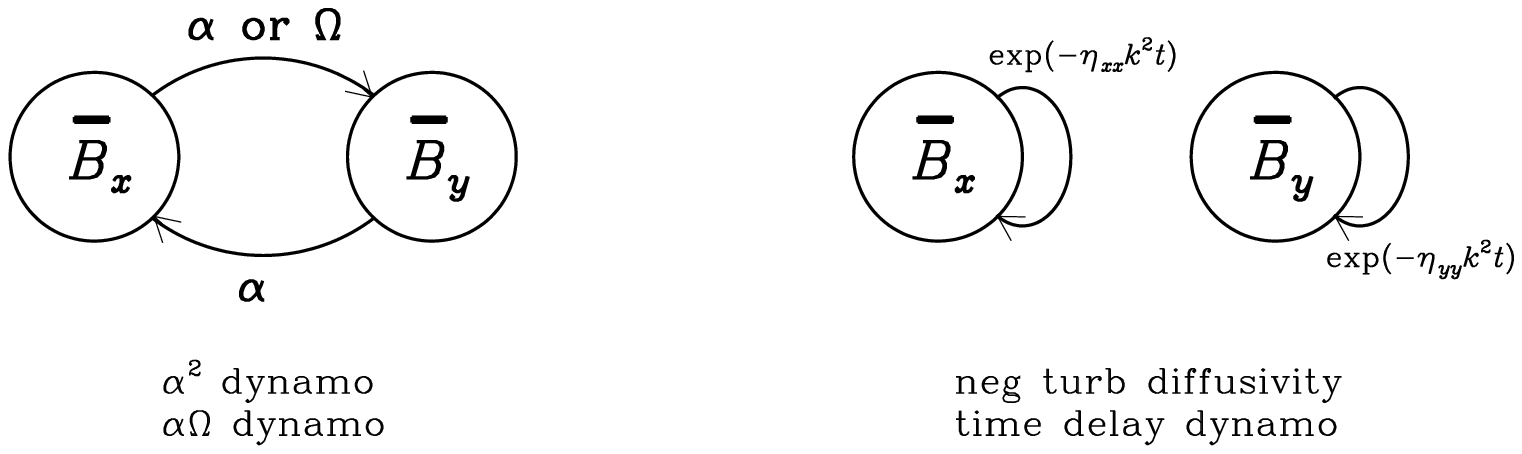}
\end{center}\caption[]{
Sketch illustrating the mutual feedbacks between $\meanB_x$ and $\meanB_y$
in $\alpha^2$ or $\alpha\Omega$ dynamos (left), and the independent
evolution of the two components in negative turbulent diffusivity and
time delay dynamos (right).
}\label{psketch}
\end{figure}

For the Willis flow, as we will see below, \Eq{dAdt} possesses solutions
with nonvanishing planar or $xy$ averages.
It turns out that there is a finite mean field, $\meanBB(z_\ast,t)$,
where $z_\ast$ is a fixed position.
In \Fig{pxyaver_bxby}, we show its $x$ and $y$ components.
Note that $\meanB_z=0$ at all times owing to the fact that
$\nab\cdot\meanBB=\partial\meanB_z/\partial z=0$ and that
the mean field was vanishing initially.
We see that $\meanB_x(z_\ast,t)$ grows with the same growth rate
as the actual magnetic field at any arbitrarily selected point
(see \Fig{pts_bxby}), but $\meanB_y(z_\ast,t)$ is seen to decay in an
oscillatory fashion with frequency $\Imag\,\lambda\approx0.58$ at a rate
$\Rey\,\lambda_{\rm mean}^{(y)}\approx-0.26$.
This is unusual and very different from the more familiar $\alpha^2$
and $\alpha\Omega$ dynamos \citep{KR80}, where poloidal and toroidal
fields always evolve in tandem.
This different behavior in comparison with standard $\alpha^2$
and $\alpha\Omega$ dynamos is illustrated in \Fig{psketch}.
Depending on the values of $\eta+\tilde{\eta}_{xx}$ and
$\eta+\tilde{\eta}_{yy}$, the $\meanB_x$ and $\meanB_y$ components
can evolve at different rates.

To investigate different planar averages in the same setup, we have
rotated the flow in both directions, $u_i(x,y,z)\to u_{i+1}(z,x,y)$
and $\to u_{i-1}(y,z,x)$, and found the same behavior in all three cases.
Below, we apply this technique to the NNT flow to study $xz$ averages
as a function of $y$.
However, to avoid confusion, we always express the final result in the
original, unrotated coordinate system.

\subsection{Dynamos in the NNT and TTT flows
and comparison with the Willis flow}

\begin{figure}\begin{center}
\includegraphics[width=\columnwidth]{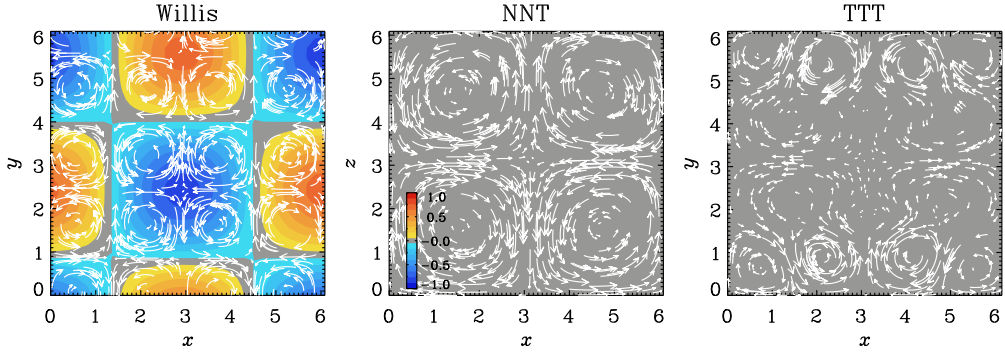}
\end{center}\caption[]{
Comparison of the three mean flows, Willis, NTT, and TTT,
obtained by averaging over the normal direction.
The flow in the plane is indicated by vectors together with the
normal component color-coded.
}\label{pum_comp}\end{figure}

The NNT flow varies very little in the $y$ direction and is dominated
by flow components in the $xz$ plane.
The TTT flow, on the other hand, varies more strongly in the $xy$ plane,
but its $z$ average still has no significant component in the $z$ direction.
For the Willis flow, the flow patterns in the $xy$, $xz$, and $yz$ planes
look the same.
In \Fig{pum_comp} we visualize $z$ averages of the Willis and TTT flows
in the $xy$ plane and $y$ averages of the NNT flow in the $xz$ plane.

\begin{figure}\begin{center}
\includegraphics[width=\columnwidth]{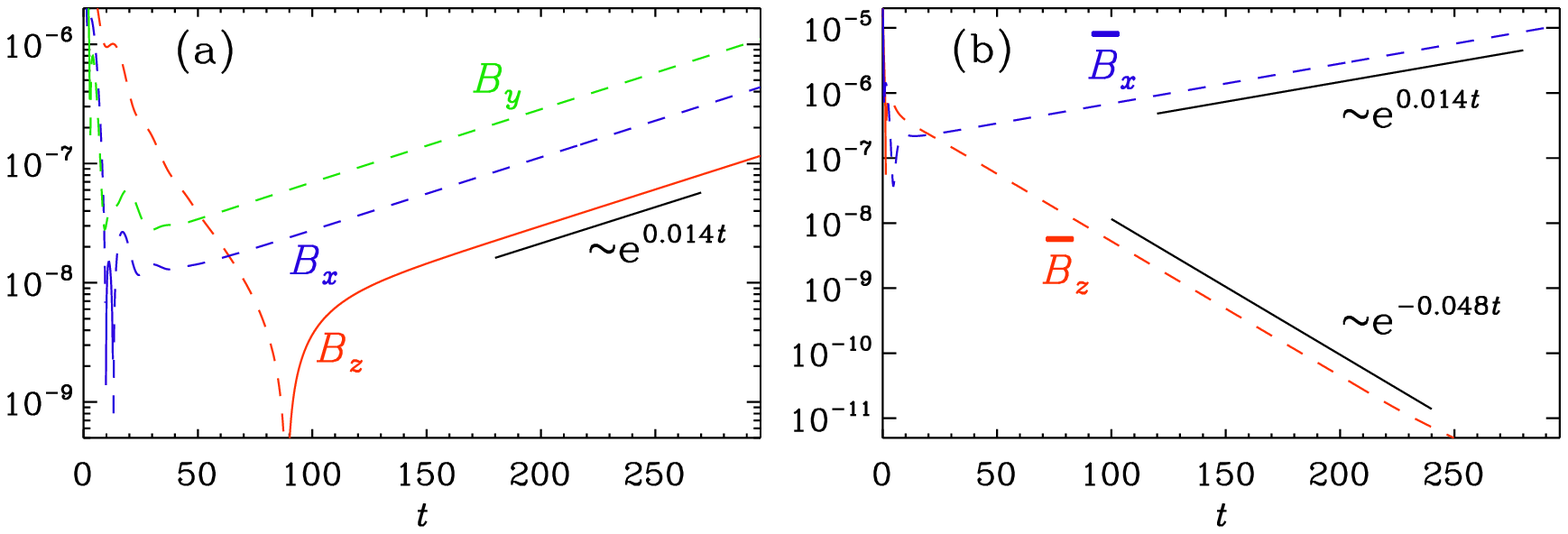}
\end{center}\caption[]{
Similar to \Figs{pts_bxby}{pxyaver_bxby}, but for the 15\% supercritical
NNT dynamo, using $xz$ planar averages.
}\label{pts_xyaver_bxby_NNT32a_rot}\end{figure}

In \Fig{pts_xyaver_bxby_NNT32a_rot}(a) we show the evolution of the
magnetic field components at selected points for the NNT flow at a
15\% supercritical value, $\eta=\eta_{\rm crit}/1.15=0.116$ and $\Rm=5.12$.
This value of $\Rm$, as defined in \Eq{rmdef},
is about $2.5$ times larger than for the 15\% supercritical case of the
Willis dynamo.
In \Fig{pts_xyaver_bxby_NNT32a_rot}(b) we show the results for $\meanB_x$
and $\meanB_y$ using $xz$ averages.
The $x$ component of the mean field grows at the same rate as the
actual field, but the $z$ component decays.

\begin{figure}\begin{center}
\includegraphics[width=\columnwidth]{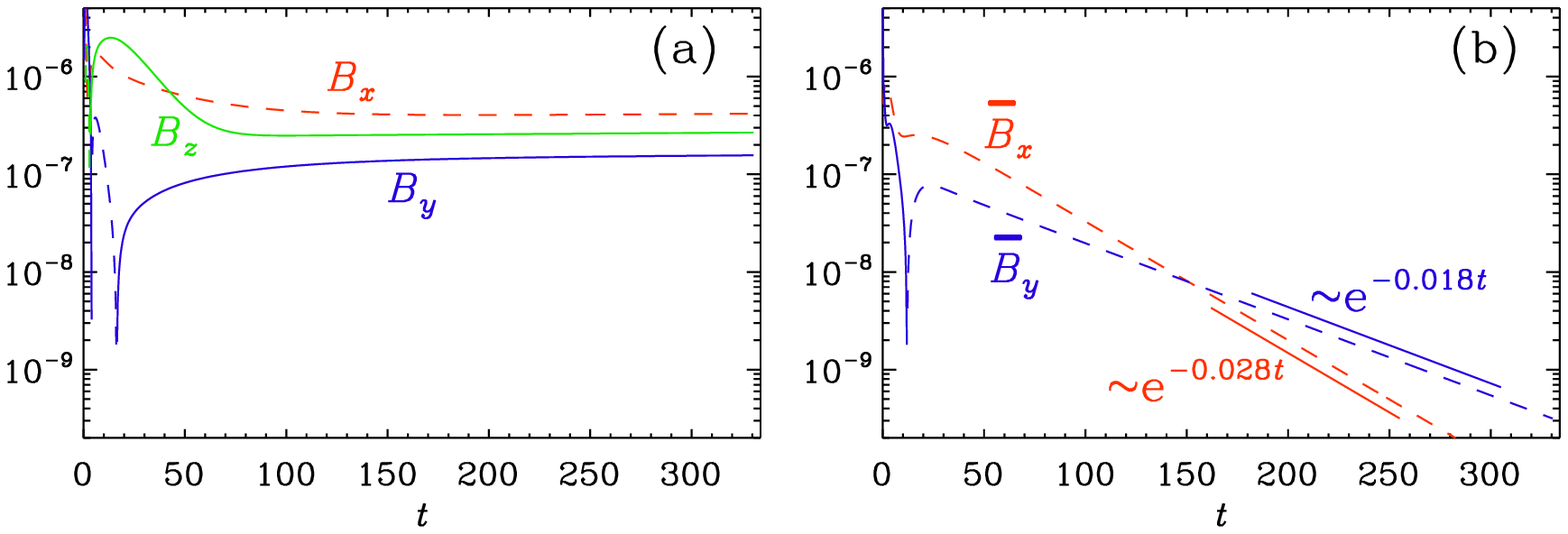}
\end{center}\caption[]{
Similar to \Fig{pts_xyaver_bxby_NNT32a_rot}, but for the marginally excited
TTT dynamo, using $xy$ planar averages. 
}\label{pts_xyaver_bxby_TTT48a}\end{figure}

For the TTT flow, we show the results for the marginally excited
case with $\eta=0.083$ in \Fig{pts_xyaver_bxby_TTT48a}.
For the TTT flow, we show in \Fig{pts_xyaver_bxby_TTT48a} the results
for the marginally excited case with $\eta=0.083$.
Here, the rms velocity is smaller than for the NNT flow,
so the magnetic Reynolds
number is actually slightly less (4.05 instead of 4.47 for the NNT case).
Furthermore, $\meanB_x$ and $\meanB_y$ now decay, both at different rates.

\begin{figure}\begin{center}
\includegraphics[width=\columnwidth]{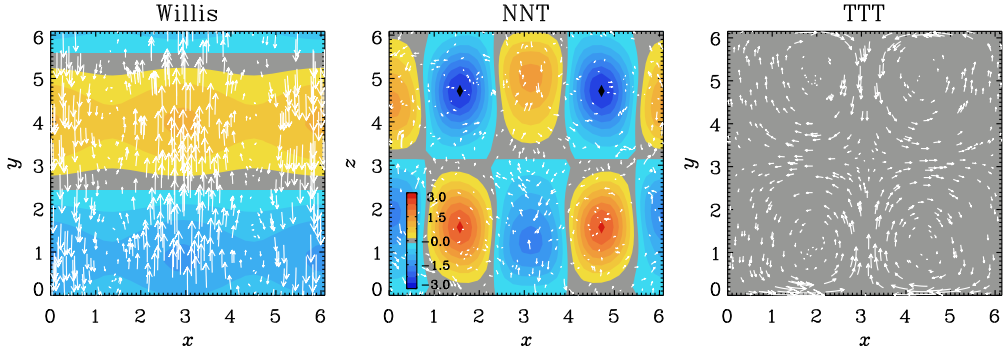}
\end{center}\caption[]{
Column averages of the magnetic fields for the Willis, NTT, and TTT
dynamos.
Similarly to \Fig{pum_comp} for the mean velocity, the mean magnetic
field in the plane is indicated by vectors together with the normal
component color-coded.
The normal component is normalized by the rms value of this mean
field based on all three components.
}\label{pbm_comp}\end{figure}

For the 15\% supercritical TTT flow, on the other hand, the actual
magnetic field is growing, but the $xy$ average remains small.
This does not necessarily imply that there is no mean field.
Indeed, a finite mean field is, in this case, obtained by taking
column averages just over the $z$ direction.
This is shown in \Fig{pbm_comp}, where we compare the resulting
column averages of the magnetic fields for the Willis, NTT, and TTT
dynamos.
We see that, even though we found a clear mean field for the Willis
flow through planar averaging, we also find a clear mean field from
just averaging over the $z$ direction.
Indeed, by computing the ratio of the rms values of the column averaged
mean field and the total field, $q=\meanBrms/\Brms$, we now find the
value 0.81 for the Willis dynamo, 0.08 for the NNT dynamo, and 0.45 for
the TTT dynamo.
The latter value was already indicated in \Tab{Tsummary} with an asterisk.
The fact that the value for the NNT flow is small suggests that the
flow does not have an important column average and that the appropriate
average is here indeed the $xy$ average.

It is important to emphasize that the Willis dynamo is completely isotropic
with respect to the $x$, $y$, and $z$ directions.
Therefore, all three planar averages give equally strong mean fields in
this linear dynamo problem.
In \Fig{pxyaver_bxby}, we plotted $\meanB_x(z_\ast)$, but this cannot be the
same mean field seen for the Willis dynamo in \Fig{pbm_comp}, which is
instead a field $\meanB_y(x)$ that is obtained through $yz$ averaging.
The third component, $\meanB_z(y)$, is also excited and can be seen
through $xz$ averaging.
It is only in the nonlinear case that one of the three planar
averages will survive, as has been demonstrated in connection with
helical isotropic turbulence; see figure~6 of \cite{Bra01}.
The other two planar averages begin to decay in the nonlinear regime.

Based on the analysis of the time series of the components of the magnetic
field, we find mean-field dynamo action for the NNT and TTT cases.
However, we still cannot say much more about the nature of the dynamo action.
To make further progress, we now use the TFM to determine the underlying
mean-field transport coefficients.
Their knowledge would allow us to model the generation of mean magnetic
fields and thereby to characterize the nature of the dynamo process.
This is discussed in the next section.
 
\section{Test-field results for the different dynamos}

\subsection{Results for the Willis flow}

\begin{figure}\begin{center}
\includegraphics[width=0.9\columnwidth]{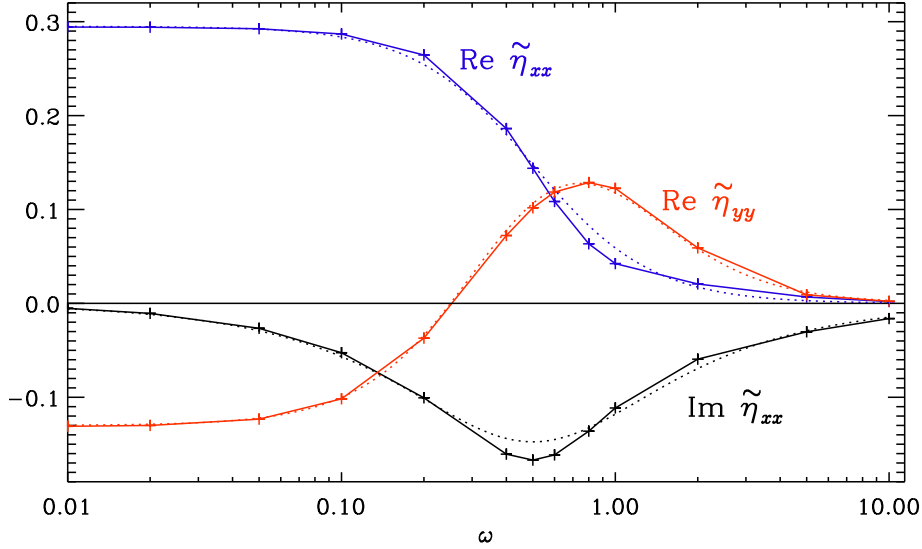}
\end{center}\caption[]{
Dependence of the real and imaginary parts of $\tilde{\eta}_{ij}$ on $\omega$
for the Willis flow in the marginally exited case with $\eta=0.403$ and for $k=1$.
The off--diagonal components vanish, $\Rey\,\tilde{\eta}_{xx}$ (blue) is always
positive, $\Imag\,\tilde{\eta}_{xx}$ (black) is always negative, and
$\tilde{\eta}_{yy}$ (red) changes sign from negative to positive values as
$\omega$ increases.
The dotted lines give approximate fits: $\tilde{\eta}_{xx}\approx0.295/(1+2\ii\omega)$
and $\tilde{\eta}_{yy}\approx0.13[-1+(4\omega)^2]/[1+(3\omega)^2+(1.6\omega)^4]$.
}\label{pk1om}\end{figure}

We begin by showing in \Fig{pk1om} the dependence of $\tilde{\eta}_{ij}$
on $\omega$ for $k=1$.
We recall that $k=1$ corresponds to the lowest wavenumber within our
domain of size $2\pi$.
We see that $\tilde{\eta}_{xx}$ and $\tilde{\eta}_{yy}$ have nonvanishing
real parts at $\omega=0$.
Furthermore, $\Rey\,\tilde{\eta}_{xx}$ is always positive and approaches zero
monotonically.
By contrast, $\Rey\,\tilde{\eta}_{yy}$ is negative, but grows and crosses zero
at $\omega=0.3$, reaches a maximum at $\omega=0.7$, and then falls off
toward zero.

\begin{figure}\begin{center}
\includegraphics[width=0.9\columnwidth]{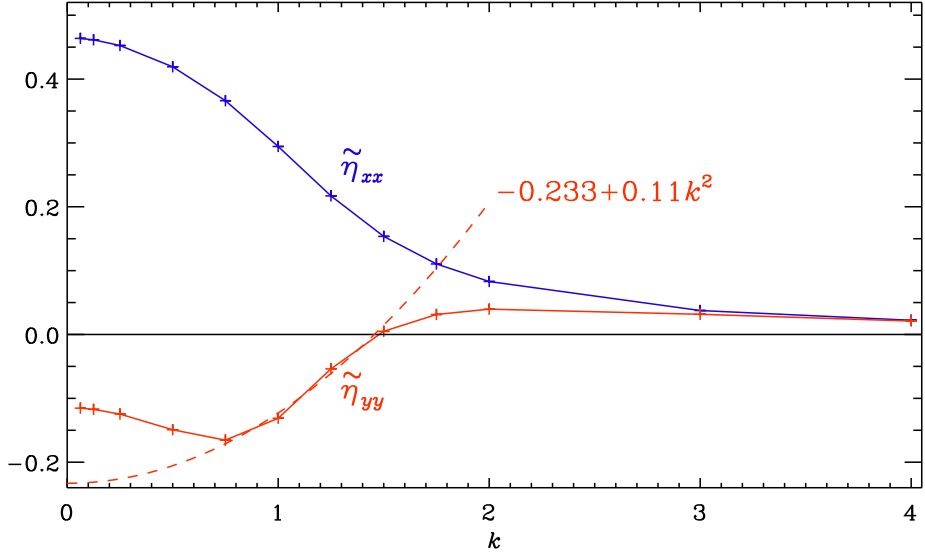}
\end{center}\caption[]{
Dependence of $\tilde{\eta}_{xx}$ (red) and $\tilde{\eta}_{yy}$ (blue) on $k$
for the Willis flow in the marginally exited case with $\eta=0.403$ and for $\omega=0$.
The dashed line denotes the fit $-0.233+0.11\,k^2$ and will be discussed
in \Sec{Negdiffdynamo}.
}\label{pom0k}\end{figure}

\begin{figure}\begin{center}
\includegraphics[width=\columnwidth]{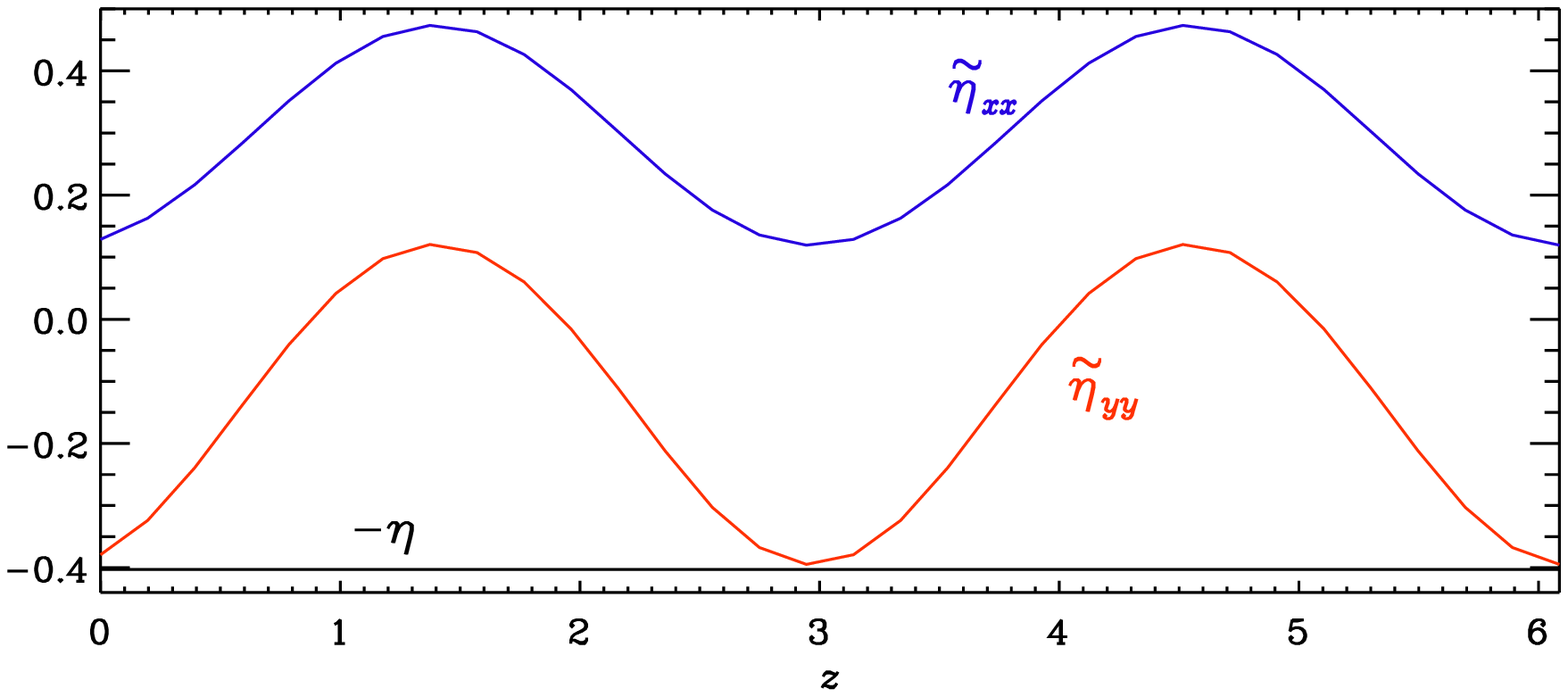}
\end{center}\caption[]{
Dependence of $\tilde{\eta}_{xx}(z)$ (blue) and $\tilde{\eta}_{yy}(z)$
(red) on $z$ for the Willis flow in the marginally exited case with
$\eta=0.4031$, $k=1$ and $\omega=0$.
}\label{pxyaver}\end{figure}

For $\omega\neq0$, $\tilde{\eta}_{xx}$ becomes complex while
$\tilde{\eta}_{yy}$ remains real.
The $\tilde{\eta}_{xx}$ component determines the evolution of $\meanA_x$ and
correspondingly $\meanB_y$, which is decaying 
even in the marginally excited case; see \Fig{pxyaver_bxby}.
The time dependence of $\meanBB$
leads a memory effect in the evolution of $\meanA_x$, i.e.,
to a frequency-dependent time delay in the electromotive force
\citep{HB09}.
By contrast, $\tilde{\eta}_{yy}$ is real for time-independent mean
magnetic fields, but it is negative for $\omega\to0$ and $k\to0$;
see \Fig{pom0k}.
However, $\tilde{\eta}_{yy}$ is not sufficiently negative to overcome
diffusive decay, because $\eta+\tilde{\eta}_{yy}$ is still positive.
This is surprising, but we have to realize that $\tilde{\eta}_{yy}$
depends also on $z$; see \Fig{pxyaver}.

\begin{figure}\begin{center}
\includegraphics[width=\columnwidth]{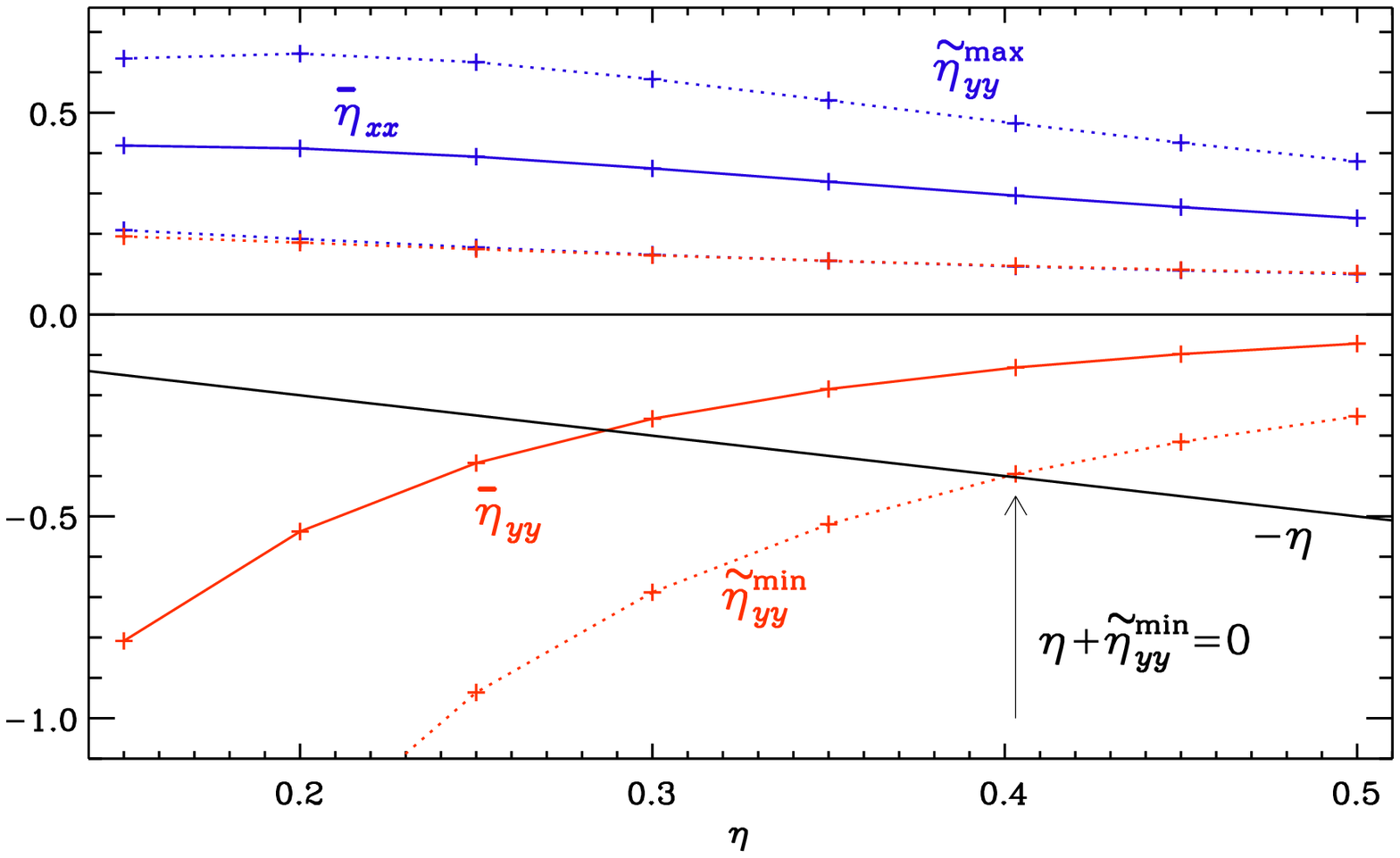}
\end{center}\caption[]{
Dependence of $\tilde{\eta}_{xx}$ (blue) and $\tilde{\eta}_{yy}$
(red) on $\eta$ for the Willis flow using $k=1$ and $\omega=0$.
The values of the minima and maxima of $\tilde{\eta}_{xx}$ and $\tilde{\eta}_{yy}$ are
shown as dotted lines.
Their $z$ averages are denoted by an overbar and are
also plotted for comparison.
The value of $-\eta$ is overplotted as a solid line to show that
$\eta+\tilde{\eta}_{yy}^{\min}=0$ when $\eta\approx0.403$. 
}\label{pk1om0eta}\end{figure}

In \Fig{pk1om0eta} we plot the local minima, maxima, and averages of
$\tilde{\eta}_{xx}$ and $\tilde{\eta}_{yy}$ versus $\eta$.
Those $z$ averages are denoted by an overbar.
We see that the onset of dynamo action ($\eta=\eta_{\rm crit}\approx0.403$)
coincides with the point where $\eta+\tilde{\eta}_{yy}^{\min}=0$.
Interestingly, it is apparently not the spatial average of
$\tilde{\eta}_{yy}(z)$ which must become negative for instability,
but the minimum of $\tilde{\eta}_{yy}(z)$.

\begin{figure}\begin{center}
\includegraphics[width=\columnwidth]{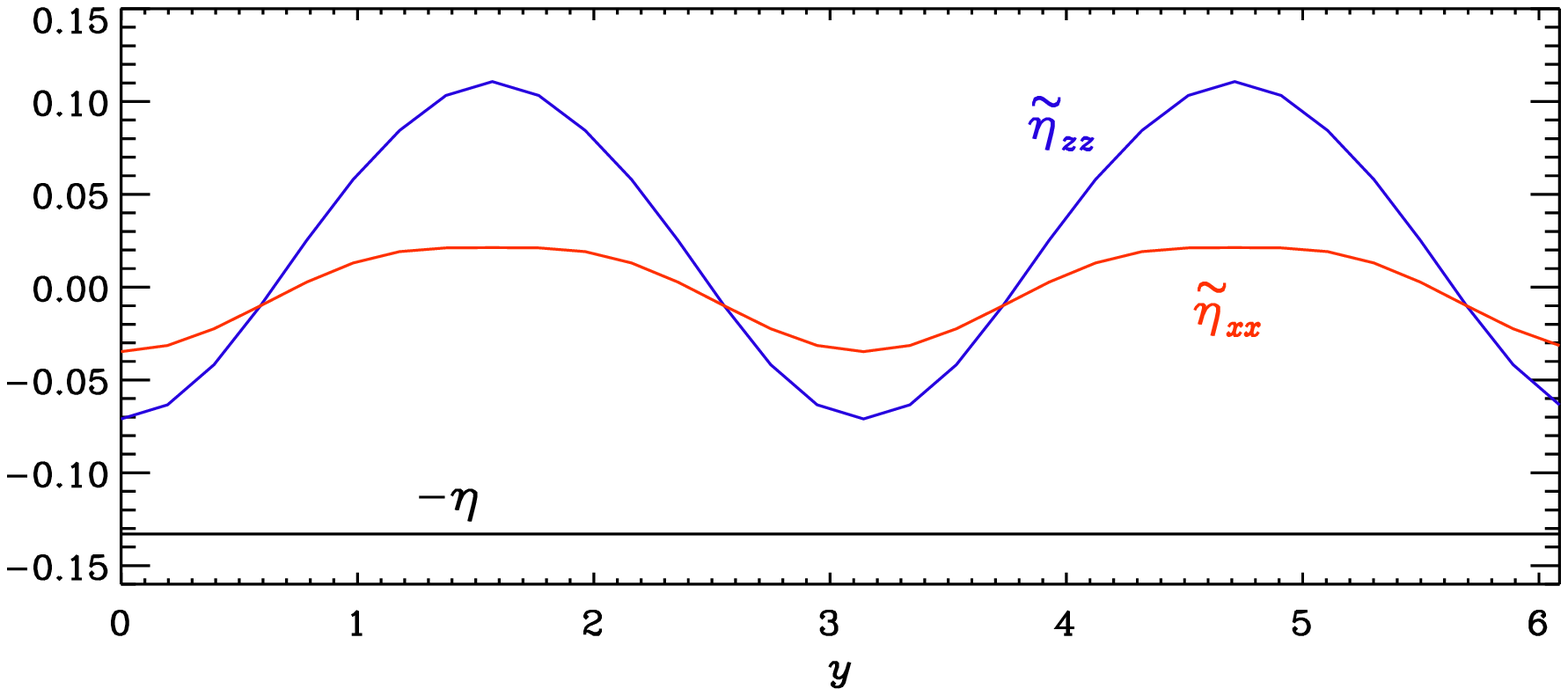}
\end{center}\caption[]{
Dependence of $\tilde{\eta}_{xx}$ (blue) and $\tilde{\eta}_{zz}$ (red)
on $y$ for the NNT flow with $\eta=0.133$, $k=1$ and $\omega=0$.
}\label{pxyaver_T32zNNTeta0133a}\end{figure}

\subsection{Results for the NNT and TTT flows}

For the NNT flow, the resulting profiles for the components of the
turbulent magnetic diffusivity are shown in \Fig{pxyaver_T32zNNTeta0133a}.
We see that both $\tilde\eta_{xx}$ and $\tilde\eta_{yy}$ become negative,
but not quite as much as to make the sum of $\eta+\tilde\eta_{ij}$
negative.
Nevertheless, it is still likely that this mean-field dynamo is due to the
negative turbulent magnetic diffusivity, because the full system is more
complicated due to the presence of pumping effects and the non-locality
in space, for example.

For the TTT case, mean-field dynamo action has only been obtained
for column averaged fields.
Applying the TFM for $xy$ averages, we find that the only nonvanishing
components of the eight transport coefficients are 
$\tilde{\eta}_{xx}$ and $\tilde{\eta}_{yy}$.
Both are positive, so no mean-field dynamo action can be expected
for $xy$ planar averages.
As we have seen above, the appropriate average is the column average.
This case is more complicated and involves altogether 27 turbulent
transport coefficients; see also \cite{War18} for a recent study in
spherical coordinates, where longitudinal averages were used.
We will not consider this case here, but adopt instead $xy$ and
$xz$ averages, as was done by \cite{A15} in the investigation of the
Taylor--Green flow, where mean fields were also only found through
column averaging.
They confirmed the presence of negative turbulent magnetic diffusivity
for this flow, which was first found by \cite{LNVW99} and was not seen
when using just planar averages \citep{DBM13}.

\begin{figure}\begin{center}
\includegraphics[width=\columnwidth]{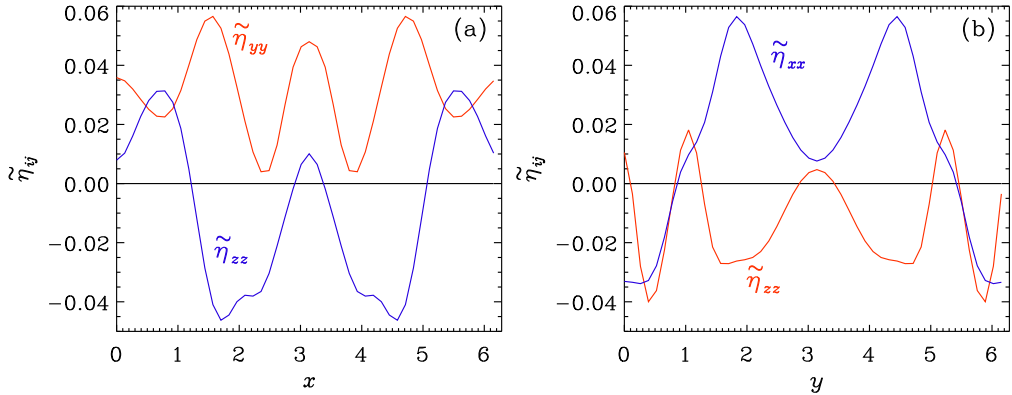}
\end{center}\caption[]{
Dependence of $\tilde{\eta}_{ij}$ 
on (a) $x$ (for $yz$ averages) and
on (b) $y$ (for $xz$ averages) for the TTT flow with $k=1$ and $\omega=0$
in a domain of size $(2\pi)^3$ using $\eta=0.083$.
}\label{pxyaver_T48TTTr}\end{figure}

In \Fig{pxyaver_T48TTTr}(a) and (b), we show the results for the TTT
flow as functions of $x$ and $y$, respectively.
Note that both $\eta_{zz}(x)$ and $\eta_{zz}(y)$ are negative over
extended ranges.
The sum $\eta+\eta_{zz}$ is still positive, but we have to remember that
the mean-field dynamo works in this case with $z$ averages, so the $yz$
and $xz$ averages adopted here are prone to additional cancellation.
This suggests that the mean-field generation in this case is indeed of
the type of a negative turbulent diffusivity dynamo.

As we will show next, for the $\pi^3$ domain, we always find a nonvanishing
average, but it may not be a mean-field dynamo, because the scales of
averaging and of the fluctuations are very similar.
To get an idea about the resulting mean-field transport coefficients,
we now solve the test-field equations in a $\pi^3$ domain
using the standard TFM for the NNT and TTT flows.
However, given that the Reynolds rules do not apply here, the
TFM results cannot be fully reliable.

\begin{figure}\begin{center}
\includegraphics[width=\columnwidth]{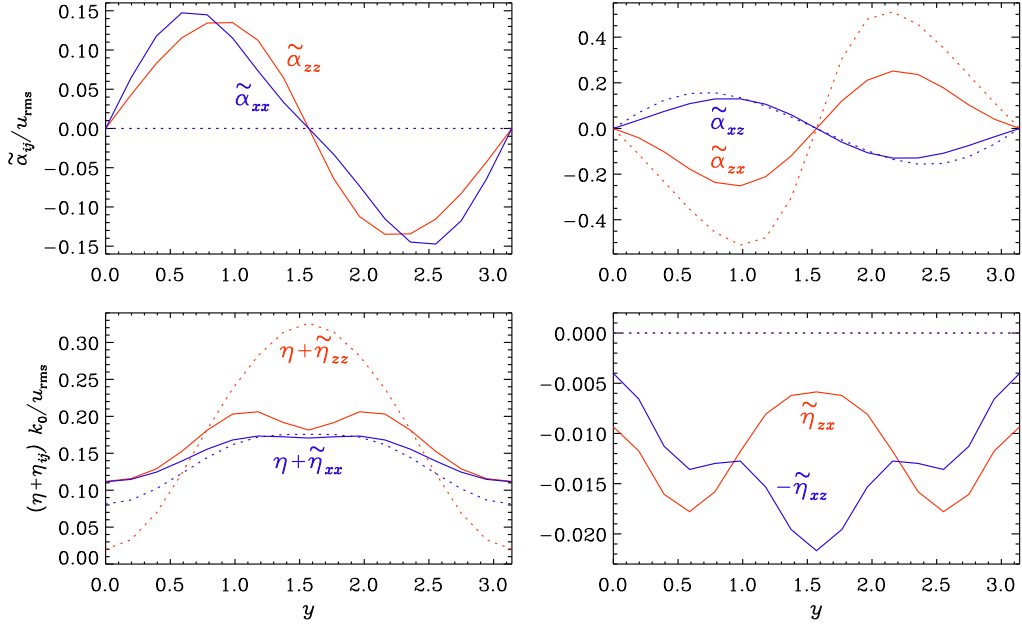}
\end{center}\caption[]{
Dependences of $\tilde{\alpha}_{ij}$ and $\tilde{\eta}_{ij}$
on $y$ for the NNT flow with $k=1$ and $\omega=0$
in a $\pi^3$ domain using $\eta=0.133$.
For the diagonal components, we show the sum $\eta+\tilde{\eta}_{ii}$.
The dotted lines denote the result using averaging over a $(2\pi)^2$ plane.
}\label{palpeta_NNT}\end{figure}

In \Figs{palpeta_NNT}{palpeta_TTT} we show test-field results for the
NNT and TTT cases in $\pi^3$ domains using the appropriate lateral and
vertical boundary conditions.
For both flows, we also have done calculations in domains where the extent
in the $y$ or $z$ directions for the NNT and TTT flows, respectively, is
from 0 to $2\pi$.
The only difference to the shorter domain is that all four components
of $\alpha_{ij}$ are antisymmetric around the middle point of the $[0, 2\pi]$
domain, while all four components of $\tilde{\eta}_{ij}$ are symmetric. 
Here we only show the results in the range $0\leq y\leq\pi$ for the
NNT flow and in the range $0\leq z\leq\pi$ for the TTT flow.

We also compare with solutions over the full $(2\pi)^2$ cross section.
The actual solutions for the response to the test fields are then
unchanged, but now the planar averages extend over the full $(2\pi)^2$
plane, so there can be complete cancellation for certain variables--- 
notably the two components of $\alpha_{ij}$ for $i=j$, as well as
for the two components of $\tilde{\eta}_{ij}$ for $i\neq j$.

\begin{figure}\begin{center}
\includegraphics[width=\columnwidth]{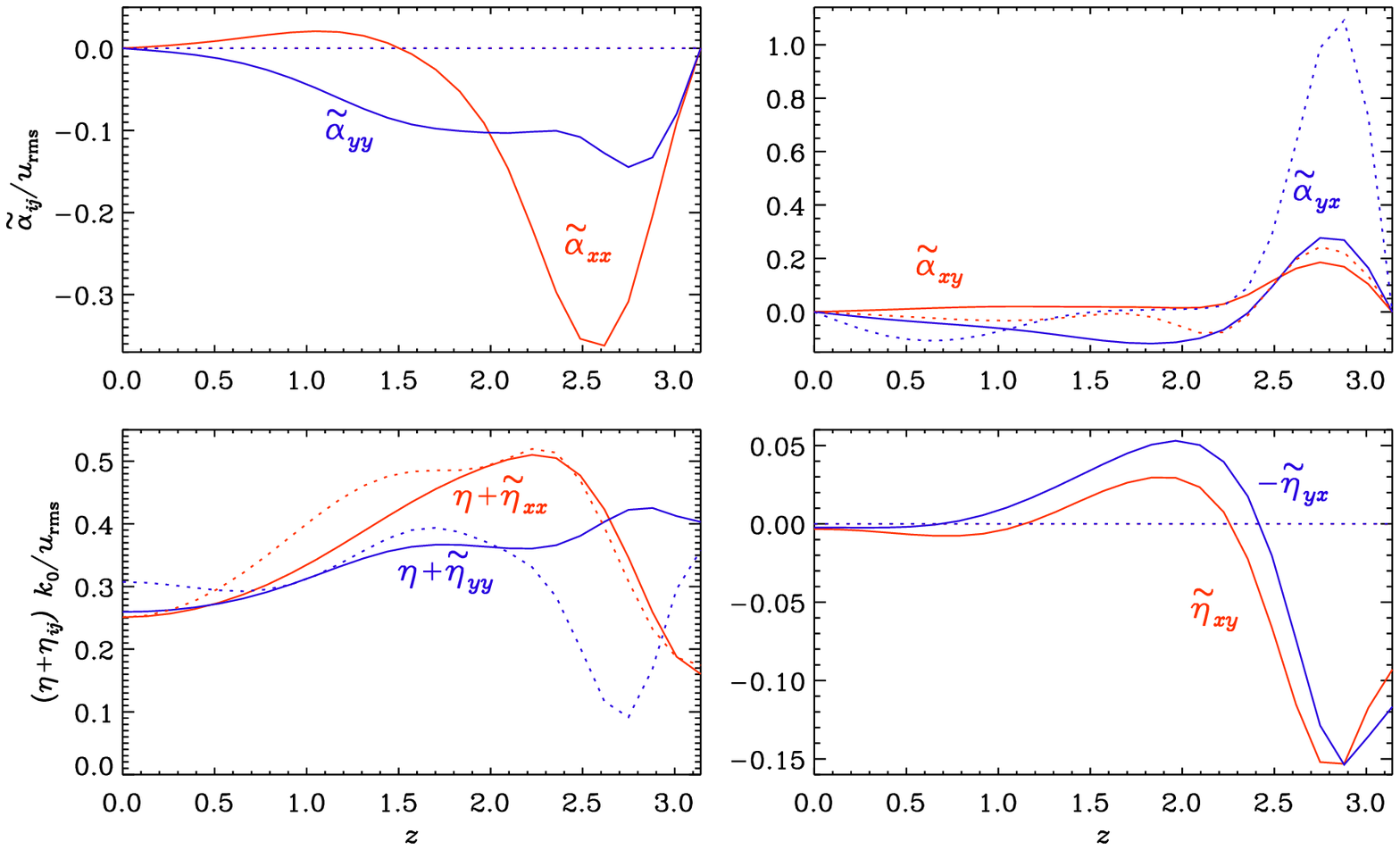}
\end{center}\caption[]{
Similar to \Fig{palpeta_NNT}, but for the TTT flow using $\eta=0.083$.
}\label{palpeta_TTT}\end{figure}

\section{Comparison with mean-field dynamos}

\subsection{The absence of $\alpha$ effect}
We now discuss in more detail the possibility of
mean-field dynamo action based on the present results.
Given that the diagonal components of $\alpha_{ij}$ are finite, there
is the possibility of $\alpha^2$ dynamo action, except that the values
of $\alpha$ appear too small in the sense that the dynamo number,
$|\bra{\alpha}|/\bra{\eta+\etat}k_1$, is below the estimated critical
value of unity.
Here, $\etat$ is the average of the two diagonal components of the
magnetic diffusivity tensor.
Furthermore, given that the off-diagonal components of
$\tilde{\eta}_{ij}$ are finite, there could be a R\"adler effect, also
known at $\OO\times\meanJJ$ effect, because the off-diagonal components
of $\tilde{\eta}_{ij}$ contribute to a term in the mean electromotive
force of the form $\meanEMF=...+\ddelta\times\meanJJ$, where $\ddelta$
is often also aligned with the axis of rotation, $\OO$.
For our planar averages, its components are given by
$\delta_i=-\half\epsilon_{ijk}\eta_{jk}$.

A well-known problem with this latter idea, however, is that the R\"adler
effect alone cannot explain an increase of the magnetic energy
of the mean field.
It can only work in conjunction with other effects such as shear.
In the presence of an $\alpha$ effect, it can also make the solutions
oscillatory and cause migratory dynamo waves, which becomes more
pronounced in a periodic domain, as can be seen by solving the
corresponding eigenvalue problem.

In the present case, we need to address the questions whether, first, the
$\alpha$ effect is strong enough to produce $\alpha^2$ or $\alpha$--shear
($\alpha\Omega$) dynamo action, depending on whether or not the mean flow plays a role,
and second, whether the R\"adler effect plays a role, possibly in
conjunction with effects other than the mean flow, for example the
$\alpha$ effect.
The first point, in fact, was already discussed by \cite{A15}.
Assuming the total magnetic field can be written as 
a series expansion 
\begin{equation}
\BB=\sum_{n=0}^\infty \bs{b}_n \epsilon^n,
\end{equation}
where $\epsilon$ is a small scaling factor that relates the 
slow-changing spatial variables to the fast-changing ones.
The $\alpha$ effect can be written 
as an eigenvalue problem for the leading term $\bs{b}_0$:
\begin{equation}
\nabla\times \bm{\alpha}\bs{b}_0=\Lambda_0 \bs{b}_0,
\end{equation}
where $\Lambda_0$ is the eigenvalue, 
$\bm{\alpha}$ is a tensor, 
$\langle \UU \times \bs{S}_i\rangle$ is the i-th column of $\bm{\alpha}$, 
$\UU$ is the flow field,
$\bs{S}_i$ is the solution of 
\begin{equation}
\mathcal{I}\bs{S}_i=-\frac{\partial \UU}{\partial x_i},
\end{equation}
and $\mathcal{I}$ is the induction operator,
\begin{equation}
\mathcal{I}\bs{b}_0=\nabla\times(\UU\times\bs{b}_0)+\eta\nabla^2\bs{b}_0.
\end{equation}
If the flow is parity invariant, 
\begin{equation}
\UU(-\bs{x})=-\UU(\bs{x}),
\end{equation}
then $\bs{S}_i$ must be anti-invariant, 
so the operator of the $\alpha$ effect is $\bm{\alpha}=0$. 
From \Eqs{willis*}{opflow}, we see that the three optimal flows 
are all parity invariant in the full $(2\pi)^3$ domain,
hence no $\alpha$ effect is expected 
at the leading order.

\subsection{The negative turbulent magnetic diffusivity dynamo}
\label{Negdiffdynamo}

To model a negative turbulent magnetic diffusivity dynamo, we must take
care of the fact that the high wavenumbers are not being destabilized
at the same time.
As we have seen from \Fig{pom0k}, $\etat$ is negative only for $k\la1.5$.
A simple way of taking the $k$ dependence of the turbulent magnetic
diffusivity into account is to expand $\tilde{\eta}_{yy}(k)$ up to the
next order in the diffusive effects (which are even in $k$), i.e.,
\begin{equation}
\tilde{\eta}_{yy}(k)=\tilde{\eta}_{yy}^{(0)}+\tilde{\eta}_{yy}^{(2)}k^2+
\ldots.
\end{equation}
Looking at \Fig{pom0k} for the Willis flow, we see that in the proximity of $k=1$,
which corresponds to the largest scale in the computational domain
of $2\pi$, the $k$-dependence of $\tilde{\eta}_{yy}(k)$ can well be
described by the parameters $\tilde{\eta}_{yy}^{(0)}\approx-0.233$
and $\tilde{\eta}_{yy}^{(2)}\approx0.11$.
In addition, there is still the microphysical magnetic diffusivity,
which is positive ($\eta=0.403$).
The mean-field equations for the magnetic field components decouple.
For the purpose of this problem, we just need to consider the equation
for $\meanA_{yy}$, which can then be written as
\begin{equation}
{\partial\meanA_{yy}\over\partial t}=
\left[\eta+\tilde{\eta}_{yy}^{(0)}\right]{\partial^2\meanA_{yy}\over\partial z^2}
-\tilde{\eta}_{yy}^{(2)}{\partial^4\meanA_{yy}\over\partial z^4}.
\label{dAmyydt}
\end{equation}
Note that the minus sign in front of the fourth derivative corresponds
to positive diffusion if $\tilde{\eta}_{yy}^{(2)}$ is positive, and so
does the plus sign in front of the second derivative, unless the term in
squared brackets is negative, which is the case we are considering here.

We have seen from \Fig{pxyaver} that $\eta+\tilde{\eta}_{yy}$ just barely
approaches zero and does not become negative.
This is unexpected.
To understand the problem, we now solve \Eq{dAmyydt} numerically with
an assumed amplitude for the variation of $\tilde{\eta}_{yy}^{(0)}$
of the form
\begin{equation}
\tilde{\eta}_{yy}^{(0)}=\overline{\eta}_{yy}^{(0)}
\left(1+\epsilon\cos2z\right),
\end{equation}
where $\epsilon$ quantifies the amplitude of the spatial variation
around the average value, which is $\overline{\eta}_{yy}^{(0)}$.
It turns out that, with the parameters given in \Fig{pom0k},
and $\eta=0.403$ for the microphysical magnetic diffusivity,
marginally excited solutions are only possible when $\epsilon>1.99$.
On the other hand, when trying to fit the functional form of
$\tilde{\eta}_{yy}^{(0)}$ in \Fig{pxyaver}, it turns out that it is
possible with $\epsilon=1.99$, but only if
$\overline{\eta}_{yy}^{(0)}=-0.133$ instead of $-0.233$, as expected
from the fit to the $k$ dependence.

It is not entirely clear how to interpret these findings.
We do not know enough about such dynamos whose parameters depend
both on $z$ and $k$ at the same time.
Nevertheless, it seems that the overall idea of a negative
turbulent magnetic diffusivity dynamo is the right one, but that
the correct description is more complicated than what is suggested
by \Eq{dAmyydt}.
Effects such as pumping and additional nonlocalities have been ignored.

\subsection{Mean-field excitation conditions for the NNT and TTT cases} 

To address the possibility of dynamo action from the $\alpha$ and
R\"adler effects, we use the profiles of $\tilde{\alpha}_{ij}(z)$ and
$\tilde{\eta}_{ij}(z)$, as obtained from the TFM.
To get an idea of how close to the dynamo onset we are, we scale the
dynamo active coefficients by an amplitude factor, i.e.,
$\tilde{\alpha}_{ij}\to\tilde{\alpha}_{ij}A_\alpha$ for $i=j$, and
$\tilde{\eta}_{ij}\to\tilde{\eta}_{ij}A_\delta$ for $i\neq j$.
We then choose a value of $A_\delta$ and determine the critical values of
$A_\alpha$ above which there is dynamo action, i.e., a growing solution.
All solutions are oscillatory and correspond to standing waves.
The result is shown in \Tab{Tsummary2n}.

\begin{table}\caption{
Critical values of $A_\alpha$ and the corresponding frequencies $\Imag\lambda$
for different values of $A_\delta$ for the NNT and TTT dynamos.
The type of averaging employed for obtaining the mean-field coefficients
is listed under ``aver'' ($xy$ or $xz$).
All solutions are standing waves.
}\vspace{12pt}\centerline{\begin{tabular}{cc|rcccccr}
flow & aver & $A_\delta$ & = & 0 & 1 & 2 & 5 & 10~~\\
\hline
NNT & $xz$ & $A_\alpha$& = & 5.00 & 4.87 & 4.77 & 4.58 & 4.63 \\
    &      & $\Imag\lambda$  & = & 0.12 & 0.13 & 0.14 & 0.17 & 0.23 \\
\hline
TTT & $xy$ & $A_\alpha$& = & 7.06 & 7.18 & 7.55 & 9.23 &11.68 \\
    &      & $\Imag\lambda$  & = & 0.08 & 0.09 & 0.07 & 0.09 & 0.17 \\
\label{Tsummary2n}\end{tabular}}\end{table}

We see that for the NNT flow, $A_\alpha$ decreases with increasing
values of $A_\delta\leq5$, so the dynamo becomes slightly easier to excite.
For $A_\delta=10$, however, $A_\alpha$ increases again.
The oscillation frequency increases slightly as $A_\delta$ increases.
For the TTT model, on the other hand, $A_\alpha$ always increases with
$A_\delta$, i.e., the $\delta$ effect does not contribute to dynamo
dynamo action, but suppresses it.
In both cases, the values of $A_\alpha$ are small unless the scaling factor $A_\delta$ 
is at least $O(10)$.

\section{Discussion and Conclusions}

There is only a small number of successful experimental dynamos to date
\citep{Riga1,Karlsruhe1,Cadarache1}.
Several other dynamo experiments are currently in operation; 
see the review by \cite{Adams}, who also discuss
the newer Madison dynamo experiment and the Derviche--Tourneur sodium
experiment.
They all work with liquid sodium, and they all involve flows with
significant swirl---in anticipation that those types of dynamos
would be most suitable for driving dynamo action most easily.
Different classes of optimal dynamos display flows largely reminiscent
of those with an $\alpha$ effect, so it was natural to ask whether such
an effect played a role in generating the magnetic field in those cases.
It now turns out that this does not have to be the case.
Instead, the optimal dynamo of \cite{W12} is driven
by a negative turbulent magnetic diffusivity.
This was rather unexpected, because such dynamos are not very common
and have not been studied much \citep{ZPF01,Zhe12,DBM13}.

The concept of negative turbulent magnetic diffusivity is somewhat
mysterious and may have seemed to be more like a qualitative excuse for lack of
a more definitive term than a quantitative and rigorous statement.
In fact, we are not aware of a quantitative mean-field model employing
negative magnetic diffusivity until now.
This would need to be done with sufficient care to prevent small-scale
instabilities.
Here we have presented such a model that is stabilized by the
inclusion of hyperdiffusivity.
In the parameter regime of interest, it turned out that the sign of the
coefficient of hyperdiffusivity is indeed such that it has a stabilizing
effect.
Among the systems with negative diffusivity and a stabilizing
hyperdiffusivity is the Kuramoto-Sivashinsky equation, which has been
employed as a model of forest fires \citep{HN77}.
In that case, however, there is also an advection term, making the
system nonlinear, and also the system size was large compared to the
typical size of structures.
This led to a rich spatio-temporal evolution.
It is therefore of interest to ask whether such dynamics could
possibly also be realized in an experimental or at least a numerical
dynamo setup.

Similarly to the Willis dynamo, the NNT dynamo also exhibits
negative turbulent magnetic diffusivity.
However, unlike the Willis dynamo, it is not isotropic and exists
only for one of the three possible planar averages.
Finally, the TTT dynamo shows no mean field after planar averaging.
Nevertheless, a mean field exists also in this case, but it requires
column averaging over only one coordinate direction.
It is similar to the Taylor--Green flow, for which \cite{LNVW99} also
found negative turbulent magnetic diffusivity, and the corresponding
mean field can only be found through column averaging \citep{A15}.
Note that both NNT and TTT flows 
can be transformed into to a general Taylor--Green flow 
by a shift of $\pi/2$ in each direction. The spatial averaging then depends 
on the specific combination of field coefficients. 

To assess the possibility of $\alpha$ or $\delta$ effect dynamos
action in $\pi^3$ subdomains, we have compared with the corresponding
mean-field dynamos.
It turned out that both are strongly subcritical and would only become
marginally excited when either the $\alpha$ effect or the $\delta$
effect are scaled up by factors between five and ten.
These factors seem rather large, making a mean-field interpretation
based on the $\alpha$ and $\delta$ effects in these cases unlikely.
Thus, it now seems that the class of dynamos based on negative turbulent
magnetic diffusivity is broader than previously anticipated and may
include the class of optimal dynamos as well.

\begin{acknowledgements}
We thank the anonymous reviewers and Phil Livermore for their useful comments and  
Maarit K\"apyl\"a, Alexei Pevtsov, Ilpo Virtanen, and Nobumitsu
Yokoi for providing a splendid atmosphere at the Nordita-supported
program on Solar Helicities in Theory and Observations. 
We would like to also thank Wietze Herreman for insightful discussions
regarding the mean-field theory.
This research was supported in part by the NSF Astronomy and Astrophysics
Grants Program (grant 1615100), the University of Colorado through
its support of the George Ellery Hale visiting faculty appointment,
and the Leverhulme Trust grant PRG-2017-169.
We acknowledge the allocation of computing resources provided by the
Swedish National Allocations Committee at the Center for Parallel
Computers at the Royal Institute of Technology in Stockholm.
\end{acknowledgements}


\end{document}